\documentclass{jpp}
\usepackage{graphicx}

\usepackage[utf8]{inputenc}
\usepackage[T1]{fontenc}
\usepackage{amsmath}
\usepackage{subfigure}
\usepackage{physics}
\usepackage{bm}
\usepackage[dvipsnames]{xcolor}
\usepackage{hyperref}

\newcommand{\vp}{v_{\parallel f}}

\newcommand{\bl}{\color{black}}
\DeclareMathOperator\erfi{erfi}

\shorttitle{Strong gradient effects on neoclassical electron transport and the bootstrap current}
\shortauthor{S. Trinczek, F. I. Parra, P. J. Catto}

\title{Strong gradient effects on neoclassical electron transport and the bootstrap current in large aspect ratio tokamaks}

\author{Silvia Trinczek\aff{1}
  \corresp{\email{strincze@pppl.gov}},
  Felix I. Parra\aff{1}
 \and Peter J.  Catto\aff{2}}

\affiliation{\aff{1}Princeton Plasma Physics Laboratory, Princeton, NJ 08543, USA
\aff{2}Plasma Science and Fusion Center, Massachusetts Institute of Technology, Cambridge, MA, USA}

\begin{document}

\maketitle

\begin{abstract}
Standard approaches to neoclassical theory do not extend into regions of strong gradients in tokamaks such as the pedestal and internal transport barriers. Here, we calculate the modifications to neoclassical electron physics inside strong gradient regions of large aspect ratio tokamaks in the banana regime. We show that these modifications are due to the different ion flow and the strong poloidal variation of the potential. We also provide a physical interpretation of the mechanisms that drive poloidal asymmetries and hence a poloidal electric field. We apply our model to two specific example cases of pedestal profiles, calculating the neoclassical electron flux and the bootstrap current. We find that depending on the ion flow, weak gradient neoclassical theory overestimates or underestimates the neoclassical electron transport and the bootstrap current in regions with strong gradients. We show that the determination of the mean parallel flow is more complex than in weak gradient neoclassical theory. For vanishing turbulence, we can determine the radial electric field for a given flow profile in the pedestal. 
\end{abstract}

\section{Introduction}
In tokamaks, strong gradients are found in the pedestal or internal transport barriers where density, temperature and the radial electric field change strongly on short length scales. Neoclassical transport can be important in these regions due to reduced turbulence levels \citep{Burrell97, Viezzer18}. One important result of neoclassical theory is the bootstrap current \citep{Bickerton71, Rosenbluth72} and its experimental validation \citep{Bonoli00, Wade04}. The bootstrap current plays a key role in macrostability as it can drive various instabilites such as the peeling-ballooning mode \citep{Connor98, Thomas04, peeters00} as well as reduce the amount of current that needs to be driven. Neoclassical theory usually assumes weak gradients \citep{Hinton76} but the bootstrap current is mainly located in the edge where gradients can be strong and this assumption is broken.\par 

\cite{Sauter99} obtained fitted expressions for the neoclassical resistivity and the bootstrap current for arbitrary aspect ratio and collisionality that were later modified by \cite{Redl21} to capture strong collisionality regimes more accurately. Since these models were fitted to results from usual neoclassical theory, it is not surprising that they have limitations in strong gradient regions, where the model by \cite{Sauter99} has been shown to overestimate the bootstrap current \citep{Hager16}. It appears that strong gradient effects indeed modify the bootstrap current.\par
These modifications in strong gradient regions have previously been considered by \cite{Kagan10b} and \cite{Shaing94, Shaing13}. In both cases, the poloidal variation of the electric potential due to strong gradient effects were not accounted for. In this work, the poloidal variation of the electric potential is kept and shown in one example to reduce the bootstrap current in the pedestal.

A neoclassical transport model for ions in strong gradient regions is presented in \cite{Trinczek23}, where the gradient length scales of density, temperature and electric potential are assumed to be of the order of the ion poloidal gyroradius, $\rho_p=\rho qR/r$. Here, $\rho$ is the ion Larmor radius, $q$ is the safety factor, $R$ is the major radius and $r$ is the minor radius. Choosing the gradient length scales to be of the order of the ion poloidal gyroradius is reasonable as this matches observations of gradient length scales in the pedestal \citep{McDermott09, viezzer2013}. Scale separation between the pedestal width and the Larmor radius $\rho$ was assumed in \cite{Trinczek23} due to an expansion in small inverse aspect ratio $r/R\sim\epsilon\ll 1$. The orbit widths of trapped and passing particles scale as $\sqrt{\epsilon} \rho_p$ and $\epsilon \rho_p$, respectively. Thus, despite keeping strong gradients, the orbit width is small and many orbits fit within one gradient length scale for $\epsilon\ll 1$. The distribution function stays close to a Maxwellian which allows an analytical treatment whilst also capturing strong gradient effects. This model includes poloidal variation, modifications to the mean parallel flow, and orbit squeezing for low collisionality. All these corrections enter as order unity modifications of the weak gradient neoclassical transport relations.\par

This article discusses strong gradient effects on neoclassical electron transport using the same framework as in \cite{Trinczek23}. The neoclassical electron transport is much smaller than the neoclassical ion transport because of the smallness of the electron-to-ion mass ratio, but the bootstrap current is sufficiently large to modify the magnetic shear and other magnetic quantities. It is to be expected that the strong gradient effects modify the bootstrap current in a similar way in which orbit squeezing, poloidal variation and modifications to the mean parallel flow modified the ion transport equations in the pedestal. We show that the poloidal variation arising from strong gradient effects in transport barriers together with the changes in the mean flow are the dominant modification mechanism of electron transport and the bootstrap current in the banana regime. The poloidal variation is caused by four different strong gradient effects: asymmetry in passing particle number, centrifugal forces, mean parallel flow gradient and asymmetry in orbit widths. The knowledge of how poloidal variation originates and how it affects neoclassical transport can be combined to study the neoclassical transport and the bootstrap current in transport barriers. \par 

The strong gradient modifications to electron neoclassical physics depend strongly on the mean parallel flow of the ions which can no longer be determined through the neoclassical ion particle flux equation. Depending on the choice of the ion parallel flow, strong gradient effects cause an increase or decrease of the bootstrap current and electron neoclassical transport in comparison to weak gradient neoclassical estimates. In this article, two example pedestal cases are presented and studied.  \par

We start in section \ref{sec:electron transport} with the derivation of the electron transport equations in the banana regime. The electron distribution function, the neoclassical electron particle flux and the bootstrap current are calculated. The poloidal variation of the electric potential enters in those transport equations. The origin of poloidal variation in strong gradient regions is discussed in more detail in section \ref{sec:poloidal}. The combination of four different strong gradient effects cause poloidal variation. This understanding is applied to two specific example cases of pedestals with different flow profiles. We find that strong gradient effects cause significant deviations from weak gradient neoclassical theory in the second example case with stronger flow gradient, but less so for the first case with weaker flow gradient. A discussion of the mean parallel flow follows in section \ref{Mean parallel flow sec}. We demonstrate that solutions to the mean parallel flow only exist for specific sources and boundary conditions if the ion neoclassical particle flux is not small. In section \ref{Radial electric field}, we study the case of a purely neoclassical pedestal without turbulence in which the ion neoclassical particle flux can be assumed to be small. For such a turbulence--free case, the transport equations in \cite{Trinczek23} can provide a solution for the radial electric field. A summary of our work and results is presented in section \ref{sec:conclusion}.

\section{Electron transport}\label{sec:electron transport}
The strong gradient modifications to the neoclassical transport of electrons are similar to those of the ion transport presented by \cite{Trinczek23}. For ions, the derivation is based on an expansion in small collisionality, assumed to be in the banana regime, $\nu_\ast\equiv qR\nu_{ee}/v_{te}\ll\epsilon^{3/2}$, and in the smallness of the inverse aspect ratio $\epsilon\ll1$. Here, $\nu_{ee}=4\sqrt{\pi} e^4 n_e\log\Lambda/(3T_e^{3/2}m_e^{1/2})$ is the electron-electron collision frequency, the electron density is denoted by $n_e$, $m_e$ is the electron mass and $\log \Lambda$ is the Coulomb logarithm and $v_{te}=\sqrt{2T_e/m_e}$ is the thermal speed of electrons with the electron temperature $T_e$. For simplicity, we work in a large aspect ratio tokamak with concentric circular flux surfaces. For electrons, the square root of the mass ratio $\delta\equiv \sqrt{m_e/m_i} \ll 1$ introduces another small parameter, where $m_i$ is the ion mass. In this work, the mass ratio and $\nu_\ast/\epsilon^{3/2}$ are the primary expansion parameters followed by an expansion in the large aspect ratio, so
\begin{equation}
   \nu_\ast/\epsilon^{3/2}\ll\epsilon\ll 1\quad \text{and}\quad \delta\ll\epsilon\ll 1.
\end{equation}
These limits are interchangeable, and starting by expanding in $\epsilon$ first would lead to the same results. \par 

The strong radial electric field introduces a shift of the trapped particle region for ions to $w\equiv v_\parallel+u\sim \sqrt{\epsilon}v_{ti}$, where $v_\parallel$ is the parallel velocity, $v_{ti}=\sqrt{2T_i/m_i}$ is the thermal speed of the ions, $T_i$ is the temperature of the ions, $u\equiv (cI/B) (\partial \Phi/\partial\psi)\sim v_{ti}$ which is related to the poloidal component of the $E\times B$-drift $\bm{v}_{E}$ via $\bm{v}_E\cdot \bnabla\theta=u \bm{\hat{b}}\cdot \nabla \theta$, $c$ is the speed of light, $\bm{B}=I\grad\zeta+\grad\zeta\times\grad\psi$ is the magnetic field, $B=\abs{\bm{B}}$ is the magnetic field strength, $\bm{\hat{b}}\equiv\bm{B}/B$ is the magnetic field direction, $\Phi$ is the electric potential, $\psi$ is the poloidal flux divided by $2\pi$, $\zeta$ is the toroidal angle, $\theta$ is the poloidal angle, and $I=RB_\zeta$ is a flux function. The shift of the trapped region introduces an asymmetry that leads to poloidal variation of density, electric potential, flow and temperature. Furthermore, the mean parallel flow is no longer set by a vanishing neoclassical ion particle flux but needs to be determined using higher order momentum conservation. The mean parallel flow profile can have a strong impact on fluxes. \par

For $T_i\sim T_e$, the shift of the trapped--particle region for electrons is small in mass ratio, $u\sim v_{ti}\sim \delta v_{te}$, where $v_{te}=\sqrt{2T_e/m_e}$ is the thermal speed of electrons, and $T_e$ is the electron temperature. The radial electric field then has a much smaller effect on electrons making it possible to neglect $u$ to lowest order in $\delta$. Thus, the condition $v_\parallel -u\sim \sqrt{\epsilon} v_{te}$ simply gives that trapped electrons have small parallel velocity $v_\parallel\sim\sqrt{\epsilon}v_{te}$. 
\par  
The main idea of our approach to calculate neoclassical electron transport is that the trapped--barely passing region has a narrow width in phase space of $v_\parallel\sim\sqrt{\epsilon}v_{te}$. Thus, most of phase space is accurately described by the freely passing particle solution and the trapped--barely passing region reduces to a discontinuity in the freely passing distribution function. It turns out that it is sufficient to calculate the height of the jump and the change in the first derivative of the passing particle distribution function across the discontinuity to derive the transport relations by integration over the drift kinetic equation. When we evaluate the transport, the height of the jump, which is set by the trapped-barely-passing particles, determines the overall flux. More details about this procedure for the ions are found in \cite{Trinczek23}. The jump contributions are derived from a drift kinetic equation which is first expanded in small collisionality. A variable transformation to so-called fixed-$\theta$ variables reduces the drift kinetic equation to a form that can be solved subsequently for the jumps across the trapped--barely passing region by an expansion in $\delta$ first and then $\epsilon$. Once the jumps have been determined, the neoclassical electron particle flux and the bootstrap current can be calculated.

\subsection{Distribution function and jump conditions}\label{sec: 2.1}
The drift kinetic equation of the distribution function $f$ can be written in the form
\begin{equation}\label{drift kinetic eq 0}
    \dot{\theta}\pdv{f}{\theta}=C[f,f]+\Sigma,
\end{equation}
where $C[f,f]$ is the collision operator and $\Sigma$ is a source term. The derivative with respect to poloidal angle $\theta$ is performed holding the magnetic moment $\mu\equiv v_\perp^2/(2B)$ and the fixed-$\theta$ variables $\vp\equiv v_\parallel(\theta_f)$ and $\psi_f\equiv \psi(\theta_f)$ fixed, where $v_\perp$ is the perpendicular speed and $\theta_f$ is a reference angle (as in \cite{Trinczek23}). To the order required $\dot\theta=(v_\parallel+u)/qR$ with $f=f(\psi_f,\theta,\vp,\mu)$. The fixed-$\theta$ variables for electrons are derived and explained in detail in Appendix \ref{Appendix fixed theta}. The source for electrons is assumed to be of order 
\begin{equation}\label{sigma}
    \Sigma_e\sim \sqrt{\epsilon}\delta^2 \nu_{ee} f_e,
\end{equation}
where $f_e$ is the electron distribution function. \par 

In the banana regime,  collisionality is small, $\nu_\ast\equiv qR\nu_{ee}/v_{te}\ll \epsilon^{3/2}$, and trapped particles complete their orbits many times before colliding. In this low collisionality limit, the drift kinetic equation to lowest order in $\nu_\ast$ is
\begin{equation}\label{no theta}
    \dot\theta\pdv{f_e}{\theta}=0.
\end{equation}
The distribution function in fixed-$\theta$ variables does not depend on poloidal angle.\par 
The transit average of \eqref{drift kinetic eq 0} eliminates the poloidal derivative and gives
\begin{equation}\label{drift kinetic eq}
    \langle C_e\rangle_\tau=-\langle \Sigma_e\rangle_\tau,
\end{equation}
where $C_e$ is the collision operator capturing electron-electron and electron-ion collisions. The transit average of a function $\mathcal{F}$  is different for trapped and passing particles. For trapped particles, it is defined as
\begin{align} \label{transit average trapped}
    \langle \mathcal{F}\rangle_\tau\equiv\frac{1}{\tau} \int_{-\theta_b}^{\theta_b}\frac{\mathrm{d}\theta}{\abs{\dot{\theta}}}\mathcal{F}(\sigma=+1)+\frac{1}{\tau}\int_{-\theta_b}^{\theta_b}\frac{\mathrm{d}\theta}{\abs{\dot{\theta}}}\mathcal{F}(\sigma=-1),
\end{align}
where
\begin{equation}
    \tau\equiv2 \int_{-\theta_b}^{\theta_b}\frac{\mathrm{d}\theta}{\abs{\dot{\theta}}},
\end{equation}
$\sigma= v_\parallel/\abs{v_\parallel}$ and $\theta_b$ is the location of the bounce point, determined to lowest order in $\delta$ by $v_\parallel=0$. It is clear from the definition \eqref{transit average trapped} that the transit average of an odd function in $\sigma$ vanishes. The transit average for passing particles is
\begin{align}
    \langle \mathcal{F}\rangle_\tau\equiv\frac{1}{\tau} \int_{-\pi}^{\pi}\frac{\mathrm{d}\theta}{\abs{\dot{\theta}}}\mathcal{F}, 
    \end{align}
    where
    \begin{align}
    \tau\equiv\int_{-\pi}^{\pi}\frac{\mathrm{d}\theta}{\abs{\dot{\theta}}}.
\end{align}
\par
The jump $\Delta \mathcal{F}$ of a function $\mathcal{F}$ will be needed later and is defined as
\begin{equation}\label{jump def}
    \Delta\mathcal{F}=\mathcal{F}^p(v_\parallel\rightarrow 0^+)-\mathcal{F}^p(v_\parallel\rightarrow 0^-)=\mathcal{F}^{bp}(w\rightarrow \infty)-\mathcal{F}^{bp}(w\rightarrow - \infty),
\end{equation}
where $\mathcal{F}^p$ and $\mathcal{F}^{bp}$ are defined in the passing and barely passing region, respectively. \par
The source term is small by $\delta^2$ according to the ordering in \eqref{sigma}. Thus, it follows from \eqref{drift kinetic eq} that $\langle C_e\rangle_\tau=0$ and hence the distribution function is an isotropic Maxwellian to lowest order in $\delta$. The electron-ion collision operator forces the electron Maxwellian to have the same flow as the ions. The ion flow is smaller than $v_{te}$ by $\delta$, and hence the Maxwellian is isotropic to lowest order in $\delta$. Equation \eqref{no theta} also imposes that $f_e$ be independent of $\theta$ when written in terms of $\psi_f$, $\vp$ and $\mu$. We choose
\begin{equation}\label{Maxwellian}
    f_e\simeq f_{Mef}=n_{e0}(\psi_f)\left(\frac{m_e}{2\pi T_{e0}(\psi_f)}\right)^{3/2}\exp\bigg\lbrace-\frac{m_e \vp^2}{2T_{e0}(\psi_f)}-\frac{m_e\mu B(\theta_f)}{T_{e0}(\psi_f)} + \frac{e\phi_1(\psi_f,\theta_f)}{T_{e0}(\psi_f)}\bigg\rbrace.
\end{equation}
Here, $\theta_f$ is a reference angle further discussed in Appendix \ref{Appendix fixed theta}. The electron density $n_{e0}$ and the temperature $T_{e0}$ in $f_{Mef}$ are only the lowest order pieces of the full electron density and temperature, defined by
\begin{align}
    n_e\equiv\int \mathrm{d}^3v\: f_e, &&  \frac{3}{2}n_eT_e\equiv\int\mathrm{d}^3v\: \frac{m_ev^2}{2}f_e.
\end{align} 
Density and temperature are hence flux functions to lowest order and can be written as
\begin{equation}
    n_e(\psi,\theta)=n_{e0}(\psi)+n_{e1}(\psi,\theta), \quad 
    T_e(\psi,\theta)=T_{e0}(\psi)+T_{e1}(\psi,\theta),
\end{equation}
where $n_{e1}/n_{e0}\sim \epsilon$ because of the Maxwell-Boltzmann response explained below, and $T_{e1}/T_{e0}\sim \delta^2$ because the first order correction to the Maxwellian will be shown to be odd in $v_\parallel$ and hence does not contribute to temperature. Similarly, the electric potential $\Phi=\phi(\psi)+\phi_1(\psi,\theta)$ has a flux function piece $\phi$ with $e\phi/T\sim 1$ and a smaller piece $\phi_1$ that depends on poloidal angle. We showed in \cite{Trinczek23} that $\phi_1/\phi\sim\epsilon$ and, for circular flux surfaces, $\phi_1(\psi,\theta)=\phi_c(\psi)\cos\theta$. The poloidally varying part of the electric potential can cause electrostatic trapping and de-trapping, thus modifying the trapping condition and number of trapped particles in the system. \par
Using equation \eqref{psi to v} and $\psi-\psi_f\sim \sqrt{\epsilon}\delta \psi\ll \psi\sim RB_p\rho_p$, where $B_p$ is the poloidal magnetic field, the distribution function can be written as
\begin{equation}\label{fe}
    f_e=f_{Mef}(\vp, \psi_f, \mu)+f_{e1f}(\vp, \psi_f, \mu)=f_{Me}(v_\parallel, \psi, \mu,\theta)+f_{e1}(v_\parallel, \psi, \mu,\theta),
\end{equation}
where
\begin{equation}\label{fMe}
    f_{Me}=n_{e0}(\psi)\left(\frac{m_e}{2\pi T_{e0}(\psi)}\right)^{3/2}\exp\bigg\lbrace-\frac{m_e v_\parallel ^2}{2T_{e0}(\psi)}-\frac{m_e\mu B(\theta)}{T_{e0}(\psi)} + \frac{e\phi_1(\psi,\theta)}{T_{e0}(\psi)}\bigg\rbrace.
\end{equation}
To lowest order, fixed-$\theta$ and particle variables are equivalent and thus the lowest order distribution function is a Maxwellian in both the fixed-$\theta$ variables and the particle variables, except for the fact that we have made it explicit in the particle variables that the density $n_e=n_{e0}(\psi)\exp\lbrace e\phi_1(\psi,\theta)/T_{e0}(\psi)\rbrace$ is not constant within flux surfaces. The relation between \eqref{Maxwellian} and \eqref{fMe} is given in Appendix \ref{Appendix lowest g} in \eqref{fme to fmef}. The correction to the Maxwellian $f_{e1}$ will be shown to have two parts. One part is of order $\delta f_{Me}$ and one part is of order $\sqrt{\epsilon}\delta f_{Me}$. An expression for $f_{e1}$ is derived in what follows. \par

The drift kinetic equation for the electrons is first expanded in $\delta$ and then in $\epsilon$. We start by expanding the collision operator. Collisions of electrons with other electrons occur as frequently as electron-ion collisions. The collision operator for electrons has to account for both electron-electron and electron-ion collisions, 
\begin{equation}\label{C electrons}
    C_e\equiv C_{ee}[f_e,f_e]+C_{ei}[f_e,f_i]\simeq C^{(l)}[f_{e1}]+\mathcal{L}\left[f_{e1}-\frac{m_e v_\parallel V_\parallel}{T_{e0}}f_{Me}\right]
\end{equation}
with $f_i$ the ion distribution function. The nonlinear terms of the collision operator are small to the order of interest and can be dropped. The self-collisions of electrons are captured by $ C^{(l)}[f_{e1}]$, which for electrons is
\begin{equation}\label{Cee}
    C^{(l)}[f_{e1}]=\bnabla_v\bcdot\left[f_{Me} \bm{\mathsf{{M_{ee}}}} \bcdot\bnabla_v\left(\frac{f_{e1}}{f_{Me}}\right)-\lambda_e f_{Me}\int\mathrm{d}^3v'\: f'_{Me} \bnabla_\omega\bnabla_\omega \omega \bcdot \bnabla_{v'}\left(\frac{f_{e1}'}{f'_{M_e}}\right)\right],
\end{equation}
where
\begin{equation}
    \bm{\mathsf{{M_{ee}}}}\equiv \lambda_e\int\mathrm{d}^3v'\: f'_{Me} \bnabla_\omega\bnabla_\omega \omega=\lambda_e\int\mathrm{d}^3v'\: f'_{Me} \frac{\omega^2\mathsf{I}-\bm{\omega}\bm{\omega}}{\omega^3},
\end{equation}
$\bm{\omega}\equiv \bm{v}-\bm{v'}$, $\lambda_e=2\pi e^4 \log\Lambda/m_e^2$ and $\bm{v}$ is the particle velocity. Collisions of electrons and ions are approximately described by a Lorentz collision operator
\begin{equation}\label{Cei}
    \mathcal{L}\left[f_{e1}-\frac{m_ev_\parallel V_\parallel}{T_{e0}}f_{Me}\right]= \bnabla_v\bcdot\left[f_{Me}\bm{\mathsf{M_{ei}}}\bcdot\bnabla_v \left(\frac{f_{e1}}{f_{Me}}-\frac{m_e v_\parallel V_\parallel}{T_{e0}}\right)\right],
\end{equation}
where
\begin{equation}
    \bm{\mathsf{{M_{ei}}}}\equiv Z^2\lambda_e n_i \frac{v^2 \mathsf{I}-\bm{v}\bm{v}}{v^3}.
\end{equation}
Here, $Z$ is the ion charge number and the ion mean parallel flow $V_\parallel$ is defined as
\begin{equation}
    n_i V_\parallel \equiv \int \mathrm{d}^3v\: v_\parallel f_i,
\end{equation}
where $n_i$ is the ion density. Just like density and temperature, the mean parallel flow has a lower order flux surface piece and a higher order piece that depends on the poloidal angle, $V_\parallel=V_{\parallel 0}(\psi)+V_{\parallel 1}(\psi,\theta)$. We separate the first order electron distribution function into two pieces,
\begin{equation}\label{fe1}
    f_{e1}=g_e+\frac{m_e v_\parallel V_\parallel}{T_{e0}}f_{Me}.
\end{equation}
The first piece will be shown to be of order $\sqrt{\epsilon}\delta f_{Me}$ and the second piece is of order $\delta f_{Me}$. With this definition, $C^{(l)}[f_{e1}]=C^{(l)}[g_e]$ because $C^{(l)}[v_\parallel f_{Me}]=0$. The combination of \eqref{Cee} and \eqref{Cei} gives \eqref{C electrons} with the final collision operator treating both electron-electron and electron-ion collisions 
\begin{equation}\label{Ce final}
    C_e= \bnabla_v\bcdot\left[f_{M_e} \bm{\mathsf{{M_{e}}}}\bcdot\bnabla_v\left(\frac{g_e}{f_{M_e}}\right)-\lambda_e f_{M_e}\int\mathrm{d}^3v'\: f'_{M_e} \bnabla_\omega\bnabla_\omega \omega \bcdot \bnabla_{v'}\left(\frac{g_e'}{f'_{M_e}}\right)\right].
\end{equation}
Here, 
\begin{equation}
 \bm{\mathsf{{M_{e}}}}\equiv\bm{\mathsf{{M_{ee}}}}+ \bm{\mathsf{{M_{ei}}}}.
\end{equation}

At this point, we can perform the same large aspect ratio expansion as for the ion calculation. We need to solve \eqref{drift kinetic eq}. In the trapped--barely passing region, $v_\parallel\sim\sqrt{\epsilon}v_{te}$ and thus the derivative of $g^{t,bp}$ with respect to parallel velocity is larger than other velocity derivatives by $\sim 1/\sqrt{\epsilon}$. Furthermore, $\nabla_v \vp\simeq v_\parallel/\vp \bm{\hat{b}}$, so we find that to lowest order, in the trapped--barely passing region, \eqref{Ce final} can be approximated by
\begin{equation}\label{Ce0}
    C_e\simeq \frac{v_\parallel}{\vp}\pdv{}{\vp}\left[\mathsf{M}_{\parallel e}\frac{v_\parallel}{\vp}\pdv{g_{e0}^{t,bp}}{\vp}\right],
\end{equation}
where 
\begin{equation}\label{Mpe}
    \mathsf{M}_{\parallel e}\equiv \mathsf{M}_{\parallel ee}+\mathsf{M}_{\parallel ei},
\end{equation}
\begin{equation}\label{Mpee}
    \mathsf{M}_{\parallel ee}\equiv \bm{\hat{b}}\bcdot \bm{\mathsf{M_{ee}}}\bcdot\bm{\hat{b}}\simeq \frac{3}{2}\sqrt{\frac{\pi }{2}}\left[\Theta(x_e)-\Psi(x_e)\right] \frac{T_e}{m_e}\frac{\nu_{ee}}{x_e} ,
\end{equation}
\begin{equation}\label{Mpei}
    \mathsf{M}_{\parallel ei}\equiv \bm{\hat{b}}\bcdot \bm{\mathsf{M_{ei}}}\bcdot\bm{\hat{b}}\simeq Z\frac{3}{2}\sqrt{\frac{\pi}{2}}\frac{T_e}{m_e}\frac{\nu_{ee}}{x_e},
\end{equation}
and $x_e^2=v^2/v^2_{te}\simeq 2\mu B/v_{te}^2$. 
In the derivation of $\mathsf{M}_{\parallel ee}$ and $\mathsf{M}_{\parallel ei}$, we have used that $u+V_{\parallel}\sim v_{ti}\ll v_{te}$ for electrons. The function $\Theta(x)=(2/\sqrt{\pi})\int_0^x\exp(-t^2)\mathrm{d}t$ is the error function and $\Psi(x)=(\Theta-x\Theta')/(2x^2)$ is the Chandrasekhar function. 
We take a transit average of \eqref{Ce0} and employ \eqref{sigma} and \eqref{drift kinetic eq} to find \bl
\begin{equation}\label{lowest trapped}
    \Bigg \langle \pdv{}{\vp}\left[ \tau \vp\left(\frac{v_\parallel}{\vp}\right)^2  \mathsf{M}_{\parallel e}\pdv {g^{t,bp}_{e0}}{\vp}\right]\Bigg\rangle_\tau=0.
\end{equation}\par
The solution to \eqref{lowest trapped} was calculated for ions by \cite{Trinczek23}. The derivation of the  electron distribution function is similar and is presented in Appendix \ref{Appendix lowest g}. The results are

\begin{align}\label{dg0e t}
    \pdv{g_{0e}^{t}}{\vp}=-\frac{\vp}{v_\parallel}\alpha_{0e}\quad\text{and}\quad \pdv{g_{0e}^{bp}}{\vp}=\left(\frac{\vp}{\langle v_\parallel\rangle_\psi}-\frac{\vp}{v_\parallel}\right)\alpha_{0e},
\end{align}
where
\begin{equation}\label{alpha0e}
   \alpha_{0e} \equiv \frac{I}{\Omega_{e}}\left[\pdv{}{\psi}\ln p_{e} +\left(\frac{m_e\mu B}{T_{e}}-\frac{5}{2}\right)\pdv{}{\psi}\ln T_{e}+\frac{m_e(u+V_{\parallel})}{T_{e}}\frac{\Omega_{e} }{I}\right]f_{Mef}
\end{equation}
The superscripts $t$ and $bp$ denote the distribution function in the trapped and the barely passing region, respectively. The electron pressure is $p_e=n_e T_e$. We can set $T_e\simeq T_{e0}$ because the difference is small in $\delta^2$, and $n_e\simeq n_{e0}$ because the difference is small in $\epsilon$. To simplify our notation, we dropped the distinction between the fixed-$\theta$ variables and ($\psi$, $v_\parallel$, $\mu$), and the difference between quantities with and without the subscripts $f$ and $0$ where possible, as these differences are small. 
Note that the electron Larmor frequency $\Omega_{e}\equiv-eB/mc$ is by definition negative and the ion Larmor frequency $\Omega_i\equiv ZeB/mc$ is by definition positive.
Integrating the expression for the electron distribution function over the trapped and barely passing region gives the height of the jump of the freely passing distribution function. The integration was carried out by \cite{Trinczek23} and gives
\begin{multline}\label{wf int}
   \Delta g_e\equiv \bigg\langle \int_{V_{t,bp}}\mathrm{d}\vp \: \pdv{g^{t,bp}_{e0}}{\vp}\bigg\rangle_\psi= \int_{V_{t,bp}}\mathrm{d}\vp \: \frac{\vp\tau}{2\pi qR}\bigg\langle \frac{v_\parallel}{\vp}\pdv{g^{t,bp}_{e0}}{\vp}\bigg\rangle_\tau \\
    = -2.758 \sqrt{\abs{\left(\mu B \frac{r}{R}+\frac{e\phi_{c}}{m_e}\right) }}\alpha_{0e},
\end{multline}
where $\langle ... \rangle_\psi\equiv 1/(2\pi)\int_{-\pi}^\pi\mathrm{d}\theta\:(...)$ is the flux surface average.
The symbol $V_{t,bp}$ denotes the trapped--barely passing region defined by $v_{\parallel f}\sim\sqrt{\epsilon}v_{te}$. The modification of the trapping condition by the poloidal variation of the electric potential results in the appearance of $\phi_{c}$ in \eqref{wf int}. The contributions from particles trapped on the low and high field side were combined by choosing first $\theta_f=0$ and then $\theta_f=\pi$ to get to the result in \eqref{wf int}.
\par

\subsection{Neoclassical electron particle flux}
Now that the jump condition \eqref{wf int} is known, we can proceed to calculate the transport relations. The electron particle flux $\Gamma_e$ is defined by the particle conservation equation,
\begin{equation}\label{particle cons}
    \pdv{\Gamma_e}{\psi_f}=\int\mathrm{d}^3 v_f\: \langle \Sigma_e\rangle_\tau.
\end{equation}
The integration over $\mathrm{d}^3v_f$ is an integration over velocity space in the fixed-$\theta$ variables, $\mathrm{d}^3v_f\equiv 2\pi B_f\mathrm{d}\mu\mathrm{d}\vp$.
Following the exact same steps as for the ion particle flux calculation, we integrate over the drift kinetic equation
\begin{equation}\label{drift e}
    -\int\mathrm{d}^3v_f \:\langle C_e\rangle_\tau=\int\mathrm{d}^3v_f \: \langle \Sigma_e\rangle_\tau.
\end{equation}
For the integration, it is useful to express the divergence in the collision operator in fixed-$\theta$ variables,
\begin{multline}\label{Ce in vf}
    \langle C_e\rangle_\tau=\\
    \frac{1}{\vp\tau}\pdv{}{\vp}\bigg[f_{Me}\vp\tau \bigg \langle \bnabla_v \vp\bcdot\Big[ \bm{\mathsf{M_{e}}}\bcdot \bnabla_v\left(\frac{g_e}{f_{Me}}\right)   -\lambda_e\int\mathrm{d}^3v'\: f'_{M_e} \bnabla_\omega\bnabla_\omega \omega \bcdot \bnabla_{v'}\left(\frac{g_e'}{f'_{M_e}}\right)\Big]\bigg\rangle_\tau\bigg]\\
    +\frac{1}{\vp\tau}\pdv{}{\mu}\bigg[f_{Me}\vp\tau \bigg \langle \bnabla_v \mu \bcdot\Big[\bm{\mathsf{M_{e}}}\bcdot \bnabla_v\left(\frac{g_e}{f_{Me}}\right)-\lambda_e\int\mathrm{d}^3v'\: f'_{M_e} \bnabla_\omega\bnabla_\omega \omega \bcdot \bnabla_{v'}\left(\frac{g_e'}{f'_{M_e}}\right)\Big]\bigg\rangle_\tau\bigg]\\
    +\frac{1}{\vp\tau}\pdv{}{\psi_f}\bigg[f_{Me}\vp\tau \bigg \langle \bnabla_v \psi_f\bcdot\Big[\bm{\mathsf{M_{e}}} \bcdot \bnabla_v\left(\frac{g_e}{f_{Me}}\right)  -\lambda_e\int\mathrm{d}^3v'\: f'_{M_e} \bnabla_\omega\bnabla_\omega \omega \bcdot \bnabla_{v'}\left(\frac{g_e'}{f'_{M_e}}\right)\Big]\bigg\rangle_\tau\bigg].
\end{multline}
The integration over the collision operator can be divided into an integration over the freely passing region and the trapped--barely passing region. Multiplying by $\vp\tau/2\pi qR$ and integrating over the freely passing region yields to lowest order in $\delta$,
\begin{equation}\label{Ce int Vp}
     -\int_{V_{p}}\mathrm{d}^3v_f\: \frac{\vp \tau}{2\pi qR}\langle C_e\rangle_\tau\simeq\int\mathrm{d}\mu\: 2\pi B_f \Delta\left[f_{M_e} \bm{\hat{b}}\bcdot \bm{\mathsf{M_{e}}} \bcdot \bnabla_v \left(\frac{g_e^p}{f_{M_e}}\right) \right]+\textit{O}(\delta^2 \epsilon^{3/2}n_e \nu_e),
\end{equation}
where we used \eqref{jump def}. Note that in the freely passing region $\vp\tau\simeq 2\pi qR$. The diffusion part of the collision operator contains the jump in the derivative of $g_e^p$ which needs to be kept. The term proportional to $\partial/\partial\mu$ in \eqref{Ce in vf} vanishes when integrating over the freely passing region. The term proportional to $\partial/\partial\psi_f$ is of order $\delta^2\epsilon^{3/2}n_e\nu_e$ because $\nabla_v \psi_f\sim\epsilon\psi_f/v_{te}$ and has been dropped. \par
There is a region of rapid $\vp$ variation for the trapped--barely passing particles. The integration gives to lowest order in $\delta$ and $\epsilon$
\begin{multline}\label{Ce int Vtbp}
     -\int_{V_{t,bp}}\mathrm{d}^3v_f\: \frac{\vp\tau}{2\pi qR}\langle C_e\rangle_\tau\simeq -\int\mathrm{d}\mu\: 2\pi B_f \Delta\left[f_{M_e} \bm{\hat{b}}\bcdot \bm{\mathsf{M_{e}}} \bcdot \bnabla_v \left(\frac{g_e^p}{f_{M_e}}\right) \right]\\
     -\pdv{}{\psi_f}\int_{V_{t,bp}}\mathrm{d}^3v_f\: \frac{\vp\tau}{2\pi qR} \frac{I}{\Omega_{e}  }\mathsf{M}_{\parallel e}\bigg\langle\left(\frac{v_\parallel}{\vp}-1\right) \frac{v_\parallel}{\vp}\pdv{g^{t,bp}_{e0}}{\vp}\bigg\rangle_\tau.
\end{multline}
In the second term, we only kept terms to order $\delta^2 \sqrt{\epsilon}n_e\nu_e$. The first term is larger than the second term by order $\delta^2$. We keep the second term in the trapped--barely passing region because the jump terms cancel when we combine \eqref{Ce int Vp} and \eqref{Ce int Vtbp},
\begin{multline}
    -\int\mathrm{d}^3v_f\: \frac{\vp \tau}{2\pi qR}\langle C_e\rangle_\tau=-\int_{V_{t,bp}}\mathrm{d}^3v_f\: \frac{\vp \tau}{2\pi qR}\langle C_e\rangle_\tau-\int_{V_{p}}\mathrm{d}^3v_f\: \frac{\vp \tau}{2\pi qR}\ \langle C_e\rangle_\tau\\
    \simeq-\pdv{}{\psi_f}\int_{V_{t,bp}}\mathrm{d}^3v_f\: \frac{\vp\tau}{2\pi qR} \frac{I}{\Omega_{e}  }\mathsf{M}_{\parallel e}\bigg\langle\left(\frac{v_\parallel}{\vp}-1\right) \frac{v_\parallel}{\vp}\pdv{g^{t,bp}_{e0}}{\vp}\bigg\rangle_\tau.
\end{multline}
The integral over $\vp$ gives the jump \eqref{wf int} such that 
\begin{equation}
     -\int_{V_{p}}\mathrm{d}^3v_f\: \frac{\vp \tau}{2\pi qR} \langle C_e\rangle_\tau =-\pdv{}{\psi_f}\left[2.758 \frac{I}{\Omega_{e} } \sqrt{\frac{r}{R}}\int\mathrm{d}\mu \: 2\pi B \mathsf{M}_{\parallel e}\sqrt{\abs{\mu B+\frac{e\phi_{c}R}{m_e r}}}\alpha_{0e} \right]
\end{equation}
since (D.16) of \cite{Trinczek23} shows $\langle(v_\parallel^2/\vp^2)\partial g_{e0}^{t,bp}/\partial\vp\rangle_\tau=0$. We can calculate the integral over $\mu$ using the expression for $\mathsf{M_{\parallel e}}$ \eqref{Mpe} and $\alpha_{0e}$ \eqref{alpha0e},
\begin{multline}\label{int for e}
    \int\mathrm{d}\mu \: 2\pi B\mathsf{M_{\parallel e}}\sqrt{\abs{\mu B+\frac{e\phi_{c}R}{m_e r}}}\alpha_{0e} =1.15 \frac{\nu_{ee}p_{e}}{m_e}\frac{I}{\Omega_{e} }\\
    \times\Bigg\lbrace \left[\pdv{}{\psi}\ln p_{e} +\frac{m_e (u+V_{\parallel })}{T_{e}}\frac{\Omega_{e} }{I}\right] G_{1e}(\phi_{c},Z)-1.39 G_{2e}(\phi_{c},Z)\pdv{}{\psi}\ln T_{e}\Bigg\rbrace.
\end{multline}
The function $G_{1e}$ is defined as
\begin{multline}\label{G1e}
    G_{1e}(\phi_{cf},Z)=\frac{\int_0^\infty \mathrm{d}x_e\: \sqrt{\abs{x_e^2+\frac{e\phi_{c} R}{T_{e} r}}}e^{-x_e^2} \left[\Theta(x_e)-\Psi(x_e)+Z\right]}{\int_0^\infty \mathrm{d}x_e\: x_e e^{-x_e^2} \left[\Theta(x_e)-\Psi(x_e)+1\right]}\\
    \simeq 1.30 \int_0^\infty \mathrm{d}x_e\: \sqrt{\abs{x_e^2+\frac{e\phi_{c} R}{T_{e} r}}}e^{-x_e^2} \left[\Theta(x_e)-\Psi(x_e)+Z\right]
\end{multline}
and $G_{2e}$ is defined as
\begin{multline}\label{G2e}
    G_{2e}(\phi_{c},Z)=\frac{\int_0^\infty \mathrm{d}x_e\:\left(x_e^2-\frac{5}{2}\right) \sqrt{\abs{x_e^2+\frac{e\phi_{c} R}{T_{e} r}}}e^{-x_e^2} \left[\Theta(x_e)-\Psi(x_e)+Z\right]}{\int_0^\infty \mathrm{d}x_e\:\left(x_e^2-\frac{5}{2}\right) x_e e^{-x_e^2} \left[\Theta(x_e)-\Psi(x_e)+1\right]}\\
    \simeq -0.94 \int_0^\infty \mathrm{d}x_e\:\left(x_e^2-\frac{5}{2}\right) \sqrt{\abs{x_e^2+\frac{e\phi_{c} R}{T_{e} r}}}e^{-x_e^2} \left[\Theta(x_e)-\Psi(x_e)+Z\right] .
\end{multline}
Combining \eqref{particle cons}, \eqref{drift e} and \eqref{int for e} gives the lowest order neoclassical electron particle flux
\begin{multline}\label{Gammae}
    \Gamma_e=-3.17\frac{\nu_{ee} I^2 p_{e}}{\Omega_{e}^2 m_e}\sqrt{\frac{r}{R}}\Bigg\lbrace \left[\pdv{}{\psi}\ln p_{e} +\frac{m_e (u+V_{\parallel })}{T_{e}}\frac{\Omega_{e } }{I}\right] G_{1e}(\phi_{c},Z)\\
    -1.39G_{2e}(\phi_{c},Z)\pdv{}{\psi}\ln T_{e}\Bigg\rbrace.
\end{multline}

The neoclassical ion particle flux for strong gradient regions from \cite{Trinczek23} is
\begin{multline}\label{Gamma}
    \Gamma_i=-1.1 \sqrt{\frac{r}{R}} \frac{\nu I^2 p_i}{\abs{S}^{3/2} m_i\Omega_i^2}\Bigg\lbrace \bigg[\pdv{}{\psi}\ln p -\frac{m_i (u+V_{\parallel})}{T_i}\left(\pdv{V_{\parallel}}{\psi}-\frac{\Omega_i}{I}\right)\bigg] G_{1}(u,V_{\parallel},\phi_c)\\
    -1.17 G_2(u,V_{\parallel},\phi_c)\pdv{}{\psi}\ln T_i\Bigg\rbrace.
\end{multline}
For ions, the particle flux depends explicitly on the mean parallel flow gradient and the squeezing factor. The functions $G_1$ and $G_2$ are defined in \citep{Trinczek23}, in (5.13) and (5.14), and depend on $u$ and $V_\parallel$, which is not the case for the electrons. The neoclassical ion and electron particle fluxes do not have to be equal. For strong gradients where $L_{n,T,\Phi}\sim \rho_p$, the fluxes are not necessarily intrinsically ambipolar \citep{sugama98, parra09, calvo12}. The neoclassical electron particle flux is then smaller than the neoclassical ion particle flux by order $\delta$. Thus, unless the turbulent particle flux compensates for the difference between $\Gamma_i$ and $\Gamma_e$, we need to impose $\Gamma_i\simeq0$. 

\subsection{The bootstrap current}\label{sec:bootstrap}
The strong gradient effects on the electrons modify the neoclassical bootstrap current $j_\parallel^B$, which is defined as 
\begin{multline}\label{jb step 1}
    j_\parallel^B\equiv Ze\int\mathrm{d}^3v\: v_\parallel f_i - e\int\mathrm{d}^3v \: v_\parallel f_e\\
    =Zen_iV_\parallel -en_eV_{\parallel } -e\int\mathrm{d}^3v\: v_\parallel g_e = -e\int\mathrm{d}^3v\: v_\parallel g_{e},
\end{multline}
where we have used quasineutrality.
The trapped--barely passing region is small in velocity space. The main contribution to the integration for the bootstrap current comes from the freely passing region where $v_\parallel=\vp+\textit{O}(\epsilon v_{te}^2/\vp)$ 
\begin{equation}\label{jb step 2}
    \langle j^B_\parallel\rangle_\psi\simeq -e\bigg\langle \int\mathrm{d}^3v \: v_\parallel g_{e} \bigg\rangle_\psi\simeq -e \bigg\langle\int_{V_p}\mathrm{d}^3v_f\: v_{\parallel } g_{e} \bigg \rangle_\psi.
\end{equation}
Here, $V_p$ denotes the freely passing region. We can calculate this integral using the Spitzer-Härm function $f_{e,SH}$ which satisfies 
\begin{equation}\label{Spitzer vparallel}
    v_\parallel f_{Me}=C_e[f_{e,SH}].
\end{equation}
The Spitzer-Härm function is a known function,
\begin{equation}\label{Spitzer function}
    f_{e,SH}=\frac{v_\parallel}{\sqrt{2}\nu_{ee}}f_{Me}A_{SH}\!\left(x_e^2\right).
\end{equation}
Here,
\begin{equation}
    A_{SH}(x_e^2)=\sum_{i} a_i L_i^{(3/2)}\left(x_e^2\right),
\end{equation}
where $L_i^{(3/2)}$ are generalized Laguerre polynomials and the coefficients $a_i$ depend on $Z$ and are tabulated. For example, the first three coefficients for $Z=1$ are $a_0=-1.975$, $a_1=0.558$ and $a_3=0.015$. 
One can use the property of self-adjointness of the collision operator in velocity space to calculate the bootstrap current. Starting with \eqref{Spitzer vparallel} inserted in \eqref{jb step 2}, self-adjointness gives 
\begin{equation}\label{self adjoint}
    \langle j_\parallel^B\rangle_\psi=-e\Bigg\langle \int\mathrm{d}^3v\: \frac{g_{e}}{f_{Me}}C_e[f_{e,SH}]\Bigg\rangle_\psi =-e\Bigg\langle \int\mathrm{d}^3v\: \frac{f_{e,SH}}{f_{Me}}C_e[g_{e}]\Bigg\rangle_\psi.
\end{equation}
We can write the expression for the bootstrap current as
\begin{equation}
    \langle j_\parallel^B\rangle_\psi\simeq-\Bigg\langle \int\mathrm{d}^3v\: \frac{e}{\sqrt{2}\nu_{ee}} v_\parallel A_{SH} C_e[g_e]\Bigg\rangle_\psi,
\end{equation}
where we used the explicit form of the Spitzer-Härm function in \eqref{Spitzer function}.
The largest contribution comes from the lowest order term in the trapped--barely passing region 
\begin{equation}
     \langle j_\parallel^B\rangle_\psi\simeq-\frac{e}{\sqrt{2}\nu_{ee}}\Bigg\langle\int_{V_{t,bp}}\mathrm{d}^3v\:  v_\parallel A_{SH} \frac{v_\parallel}{\vp}\pdv{}{\vp}\left[ \mathsf{M}_{\parallel e} \frac{v_\parallel}{\vp}\pdv{g_{0e}^{t,bp}}{\vp}\right]\Bigg\rangle_\psi\sim \sqrt{\epsilon}\delta n_ie v_{te}.
\end{equation}
Using $\mathrm{d}^3v=(\vp/v_\parallel)\mathrm{d}^3v_f$, we can make a change to fixed-$\theta$ variables, 
\begin{multline}
     \langle j_\parallel^B\rangle_\psi\simeq-\frac{e}{\sqrt{2}\nu_{ee}}\Bigg\langle\int_{V_{t,bp}}\mathrm{d}^3v_f\: v_\parallel  A_{SH} \pdv{}{\vp}\left[ \mathsf{M}_{\parallel e} \frac{v_\parallel}{\vp} \pdv{g_{0e}^{t,bp}}{\vp}\right]\Bigg\rangle_\psi\\
     =\frac{e}{\sqrt{2}\nu_{ee}}\Bigg\langle\int_{V_{t,bp}}\mathrm{d}^3v_f\: A_{SH}\mathsf{M}_{\parallel e} \pdv{g_{0e}^{t,bp}}{\vp}\Bigg\rangle_\psi,
\end{multline}
where we integrated by parts in the second step and we employed $\partial v_\parallel/\partial\vp=\vp/v_\parallel$. For trapped and barely passing particles, $m_ev^2/(2T_e)\simeq m_e\mu B/T_e$, so $A_{SH}$ is independent of $\vp$. The integration over $\vp$ gives the jump $\Delta g_e$, which is given in \eqref{wf int}, and we arrive at,
\begin{equation}\label{bootstrap integral}
     \langle j^B_\parallel\rangle_\psi=-2.758\sqrt{\frac{r}{R}}\frac{e}{\sqrt{2}\nu_{ee}}2\pi B \int\mathrm{d}\mu\: \sqrt{\abs{\mu B +\frac{e\phi_{c} R}{m_e r}}}\alpha_{0e}  \mathsf{M_{\parallel e}}A_{SH}\left(\frac{m_e\mu B}{T_e}\right).
\end{equation}
The expression for $\alpha_{0e}$ is given in \eqref{alpha0e}. Appendix \ref{Appendix bootstrap} gives a derivation that treats the discontinuities more carefully but demonstrates that our procedure presented here is completely consistent with the jump conditions that we calculated in section \ref{sec: 2.1}.  
\par The neoclassical bootstrap current including strong gradient effects is 
\begin{multline}\label{jb}
    \langle j_\parallel ^B\rangle_\psi=-2.43 \frac{cIp_{e}}{B}\sqrt{\frac{r}{R}}\bigg[\left(\pdv{}{\psi} \ln p_{e}+\frac{m_e(u+V_{\parallel })}{T_{e}}\frac{\Omega_{e} }{I}\right) J_{1e}(\phi_{c},Z)\\
    -0.71J_{2e}(\phi_{c},Z)\pdv{}{\psi}\ln T_{e} \bigg],
\end{multline}
where
\begin{multline}\label{J1}
    J_{1e}(\phi_c,Z)=\frac{\int_0^\infty\mathrm{d}x\: \sum_{i} a_iL_i^{3/2}(x^2)\left[ \Theta(x)-\Psi(x) +Z\right]e^{-x^2}\sqrt{\abs{x^2+\frac{e\phi_{c} R}{T_{e} r}}}}{\int_0^\infty\mathrm{d}x\: \sum_{i} a_{i,Z=1}L_i^{3/2}(x^2)\left[ \Theta(x)-\Psi(x) +1\right]e^{-x^2}x}\\
    \simeq -1.2 \int_0^\infty\mathrm{d}x\: \sum_{i} a_iL_i^{3/2}(x^2)\left[ \Theta(x)-\Psi(x) +Z\right]e^{-x^2}\sqrt{\abs{x^2+\frac{e\phi_{c} R}{T_{e} r}}}
\end{multline}
and
\begin{multline}\label{J2}
    J_{2e}(\phi_c,Z)=\frac{\int_0^\infty\mathrm{d}x\:\left(x^2-\frac{5}{2}\right) \sum_{i} a_iL_i^{3/2}(x^2)\left[ \Theta(x)-\Psi(x) +Z\right]e^{-x^2}\sqrt{\abs{x^2+\frac{e\phi_{c} R}{T_{e} r}}}}{\int_0^\infty\mathrm{d}x\: \left(x^2-\frac{5}{2}\right)\sum_{i} a_{i,Z=1}L_i^{3/2}(x^2)\left[ \Theta(x)-\Psi(x) +1\right]e^{-x^2}x}\\
    \simeq 1.7 \int_0^\infty\mathrm{d}x\: \sum_{i} a_iL_i^{3/2}(x^2)\left[ \Theta(x)-\Psi(x) +Z\right]e^{-x^2}\sqrt{\abs{x^2+\frac{e\phi_{c} R}{T_{e} r}}}.
\end{multline}

The weak gradient expression for the bootstrap current is modified by the poloidal variation of the electric potential, which is captured by the modification of the coefficients via the functions $J_{1e}$ and $J_{2e}$. The bootstrap current also depends on the ion flow which can be different in the strong gradient region as it is no longer determined through flow damping, see \cite{Trinczek23}. The origin of the poloidal variation and its effects on the electron particle flux, the bootstrap current and the mean parallel flow are further discussed in the next sections.

\section{Poloidal variation}\label{sec:poloidal}
Both the electron particle flux and the bootstrap current are modified with respect to the usual neoclassical expressions via the coefficients $G_{1e}$, $G_{2e}$, $J_{1e}$, and $J_{2e}$, which are functions of the amplitude of the poloidally varying part of the potential, $\phi_1=\phi_c(\psi)\cos\theta$. 
The possibility of poloidal variation of the electric potential modifying neoclassical transport and the bootstrap current was already considered by \cite{Chang83}, although he did not calculate the poloidally varying part of the electric potential. Impurity measurements of H-mode pedestals on Alcator C-Mod \citep{theiler14, churchill15} and Asdex-Upgrade \citep{Cruz22} have demonstrated poloidal asymmetry in density, electric field and ion temperature. \cite{Trinczek23} found that neoclassical effects in regions with large gradients can produce poloidal asymmetries similar to the ones measured in pedestals. Impurity injection is also responsible for poloidal variation \citep{Helander98}. At large aspect ratios, the model by \cite{bielajew23} allows $e\phi_1/T\sim\epsilon$ (with up-down as well as in-out asymmetries) but it cannot treat strong gradients since it assumes $e\phi/T\sim \epsilon$ and is thus not applicable in strong gradient regions at present. 
Here, we combine the poloidal variation calculated by \cite{Trinczek23} with our formulas for electron physics. First we revisit the origin of the in-out poloidal variation and complete the physical picture in \cite{Trinczek23}, then we apply the transport calculation to a specific set of pedestal profiles to understand how the strong gradient effects act through poloidal variation.

\subsection{Origin of poloidal variation}\label{sec: phic effects}
The amplitude $\phi_c$ of the part of the electric potential that depends on poloidal angle was derived by \cite{Trinczek23}. The final result reads
\begin{equation} \label{phic}
    \begin{split}
        &\Bigg\lbrace\frac{en_e}{T_e}\!-\!\frac{Z^2n_ie I}{T_i\Omega_i}\!\left[\!\sqrt{\frac{2T_i}{m_i}}\!J\!\left(\!\pdv{}{\psi}\ln p_i\!-\!\frac{3}{2}\pdv{}{\psi}\ln T_i\right)\!+\!\left[1\!-\!2\sqrt{\frac{m_i}{2T_i}}(V_\parallel+u)J\right]\right(\pdv{V_\parallel}{\psi}\!-\!\frac{\Omega_i}{I}\!\\
        &-\!\frac{(V_\parallel+u)}{2}\pdv{}{\psi}\ln T_i\Bigg)\Bigg]\Bigg\rbrace\phi_c=-Zn_i\frac{Ir}{\Omega_i R}\Bigg\lbrace \sqrt{\frac{2T_i}{m_i}}J \Bigg[\left(\frac{m_iV_\parallel^2}{T_i}+1\right)\left(\pdv{}{\psi}\ln p_i-\frac{3}{2}\pdv{}{\psi}\ln T_i\right)\\
        &+\pdv{}{\psi}\ln T_i\Bigg]+\!\left[1\!-\!2\sqrt{\frac{m_i}{2T_i}}(V_\parallel+u)J\right]\!\Bigg[\!(V_\parallel\!-\!u)\!\left(\pdv{}{\psi}\!\ln \!p_i\!-\!\frac{3}{2}\pdv{}{\psi}\!\ln\! T_i\right)\!\\
        &+\!\left(\pdv{V_\parallel}{\psi}\!-\!\frac{\Omega_i}{I}\right)\!\left(\frac{m_iu^2}{T_i}\!+\!1\!-\!\frac{m_i(V_\parallel+u)^2 }{T_i}\right)\!-\frac{V_\parallel+u}{2}\left(\frac{m_iV_\parallel^2}{T_i}+1\right)\pdv{}{\psi}\ln T_i\Bigg]\\
        &+\left[1+2\frac{m_i}{2T_i}(V_\parallel+u)^2-4\left(\frac{m_i}{2T_i}\right)^{3/2}(V_\parallel+u)^3J\right]\left(\pdv{V_\parallel}{\psi}-\frac{\Omega_i}{I}+\frac{V_\parallel-u}{2}\pdv{}{\psi}\ln T_i\right)\Bigg\rbrace\\
        &-2Zn_i\frac{r}{R},
    \end{split}
\end{equation}
where
\begin{equation}
    J\equiv\frac{\sqrt{\pi}}{2}\exp\left[-\frac{m(u+V_\parallel)^2}{2T}\right]\erfi\left[{\sqrt{\frac{m}{2T}}(u+V_\parallel)}\right]
\end{equation}
and $\erfi(x)\equiv(2/\sqrt{\pi})\int_0^x\exp(t^2)\: \mathrm{d}t$. The strong gradients cause poloidal variation in four different ways: passing particle number asymmetry, centrifugal forces, orbit width asymmetry and mean parallel flow gradient. The first effect, passing particle number asymmetry, was presented in detail in section 4.4 and figure 5 in \cite{Trinczek23}. In this paper we want to explain the other three strong gradient effects that cause poloidal variation and were not explicitly mentioned in \cite{Trinczek23}. \par
We start by reminding the reader that there is a passing particle number asymmetry, and that this asymmetry causes poloidal variation in the density, flow, temperature and potential. For $V_\parallel\neq-u$, the passing particle region is no longer symmetric around the trapped particle region. This causes an asymmetry in the number of passing particles circulating in the positive and negative poloidal direction. For example, for $V_\parallel>-u$ more particles circulate poloidally in the positive direction than in the negative one. For any flux surface of interest, there are two groups of particles at the outboard and at the inboard side: one group with positive and one with negative poloidal velocity. The average radial position of the particles with positive poloidal velocity on the outboard side lies inside the flux surface of interest, that is, in the high density region, whereas the average radial position of the other group of particles, the group with negative poloidal velocity on the outboard side, is in the region of slightly lower density. Due to the asymmetry in passing particle numbers, this creates a point of slightly higher density on the outboard sign. For the inboard side, this picture reverses and a point of slightly lower density is created.\par
Setting $V_\parallel+u=0$ eliminates the asymmetry in passing particle number, yet the poloidal variation of the electric potential does not vanish, 
\begin{multline} \label{phicterms}
        \left[\frac{en_e}{T_e}-\frac{Z^2n_ie I}{T_i\Omega_i}\left(\pdv{V_\parallel}{\psi}-\frac{\Omega_i}{I}\right)\right]\phi_c= -Zn_i\frac{Ir}{\Omega_i R}\bigg[-\frac{\Omega_i}{I}\frac{m_i V_\parallel^2}{T_i}+2\pdv{V_\parallel}{\psi}\left(1+\frac{m_iV_\parallel ^2}{2T_i}\right)\\
        +2V_\parallel\pdv{}{\psi}\ln n_i\bigg].
\end{multline}

\par The first term which is proportional to $m_iV_\parallel^2/T_i$ is the centrifugal force. The centrifugal force pushes ions to the outboard side. The electrons are lighter and less affected by the centrifugal force, so an electrostatic potential is created that is positive on the outboard side and negative on the inboard side to ensure quasineutrality. This effect vanishes in the low flow limit of weak gradient theory because $V_\parallel$ is small.\par

\begin{figure}
\centering
    \subfigure[]{ \includegraphics[width=0.35\textwidth]{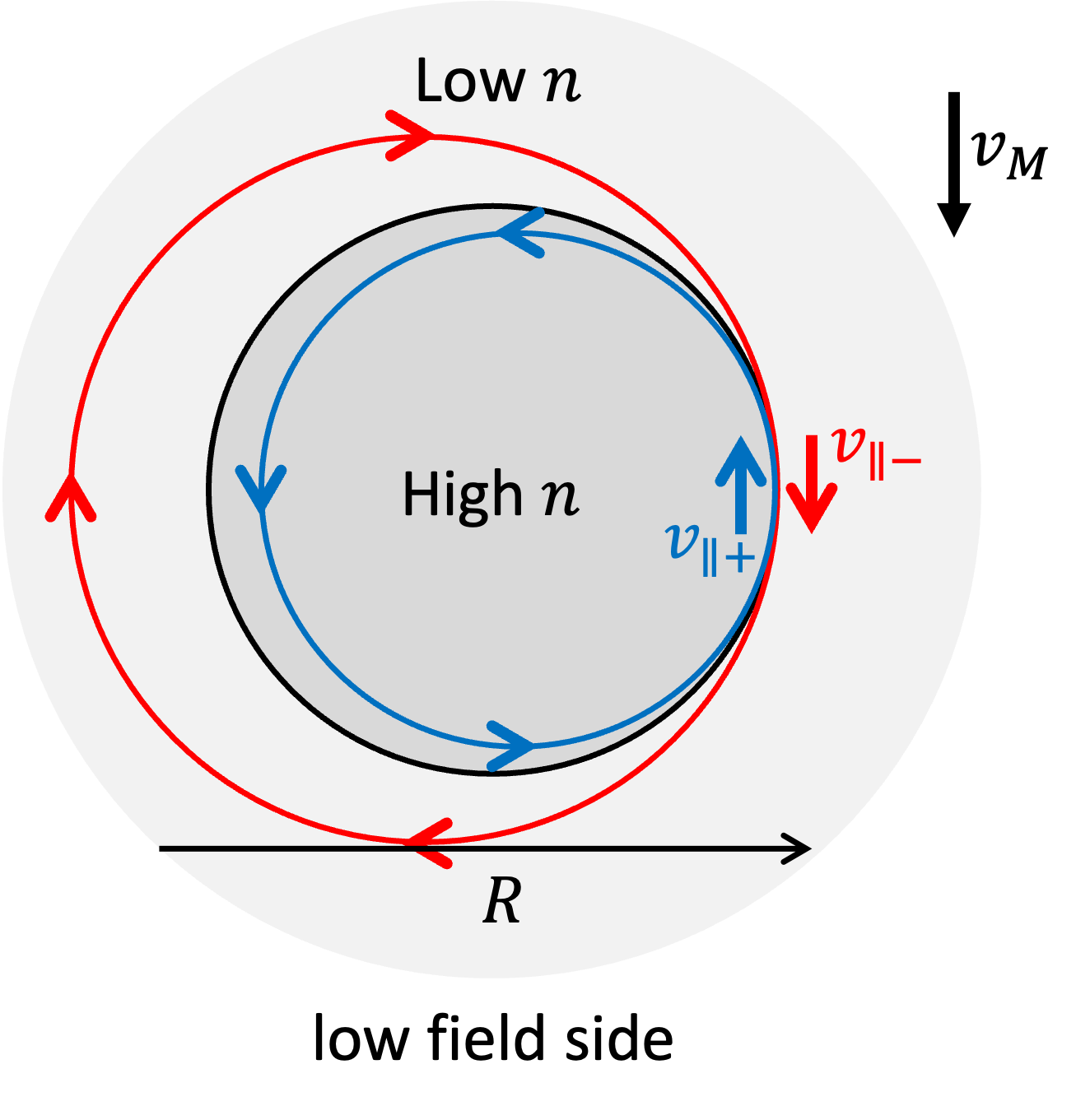}}
    \subfigure[]{ \includegraphics[width=0.32\textwidth]{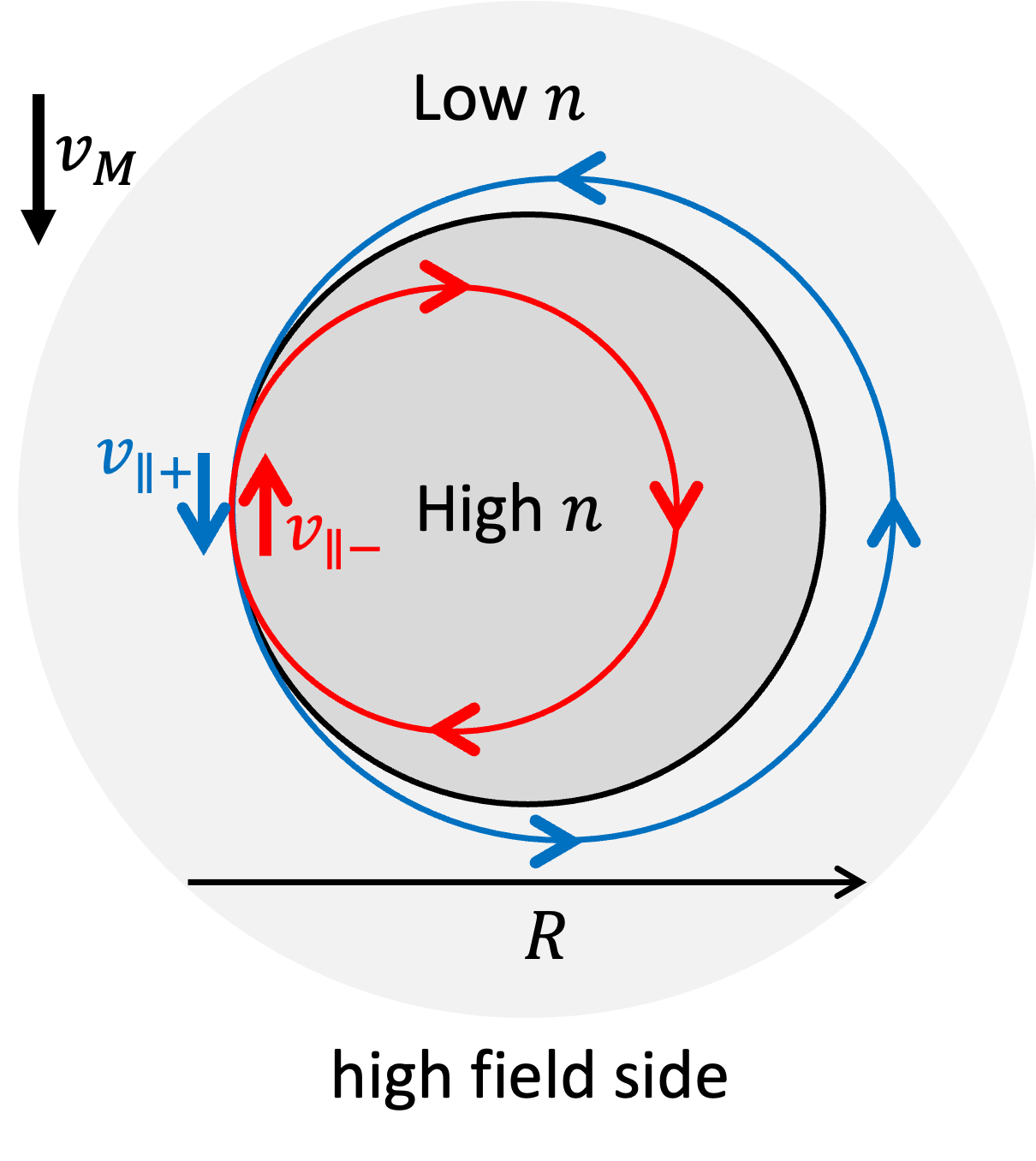}}
    \caption{The orbit width of co- and counter-circulating particles is asymmetric because of curvature drift. In this figure, we assume $V_\parallel=-u<0$. On the low field side, red particles have a larger orbit width so their average radial position locates them deeper in the low density region than blue particles are located in the high density region (a). The opposite happens on the high field side (b). This creates a higher density on the high field side than on the low field side. This effect depends on the sign of $V_\parallel$ and reverses for positive $V_\parallel$. }
    \label{fig: widths}
\end{figure}

The last term is proportional to the density gradient and $V_\parallel$. This term is related to the asymmetry in orbit widths. Looking back at the ion orbit equations for passing particles derived by \cite{Trinczek23}, particles with negative poloidal velocity have a slightly larger orbit width than particles with positive poloidal velocity, that is, the orbit widths are not symmetric in $v_\parallel+u$. The asymmetry is caused by the curvature drift which is symmetric in $v_\parallel$ but not with respect to $v_\parallel+u=0$. We explain this effect in figure \ref{fig: widths}, where we assume $u>0$ and hence $V_\parallel<0$. Particles with parallel velocity $v_{\parallel -}<V_\parallel=-u$ in figure \ref{fig: widths} experience a stronger curvature drift than particles with parallel velocity $v_{\parallel +}>V_\parallel=-u$. In other words, the red particles have a larger orbit width and move away from their flux surface further than the blue particles, see figure \ref{fig: widths}. On the outboard side, the average radial position of the red particles is deeper into the low density region than the average radial position of the blue particles is in the high density region. The outboard side turns into a point of slightly lower density. The picture again reverses for the low field side, where the inward going particles travel further in radius such that the high field side has slightly higher density than the low field side and poloidal variation occurs. This effect depends on the sign of $V_\parallel$. In this argument and in figure \ref{fig: widths}, we assumed that $u>0$ and thus $V_\parallel=-u<0$. If $V_\parallel>0$, $\abs{v_{\parallel-}}<\abs{v_{\parallel+}}$, so the orbit width of the blue particles would be bigger. In the limit of weak gradients, this effect vanishes because $V_\parallel $ and the density gradient are small.\par

\begin{figure}
    \centering
   \subfigure[]{ \includegraphics[width=0.48\textwidth]{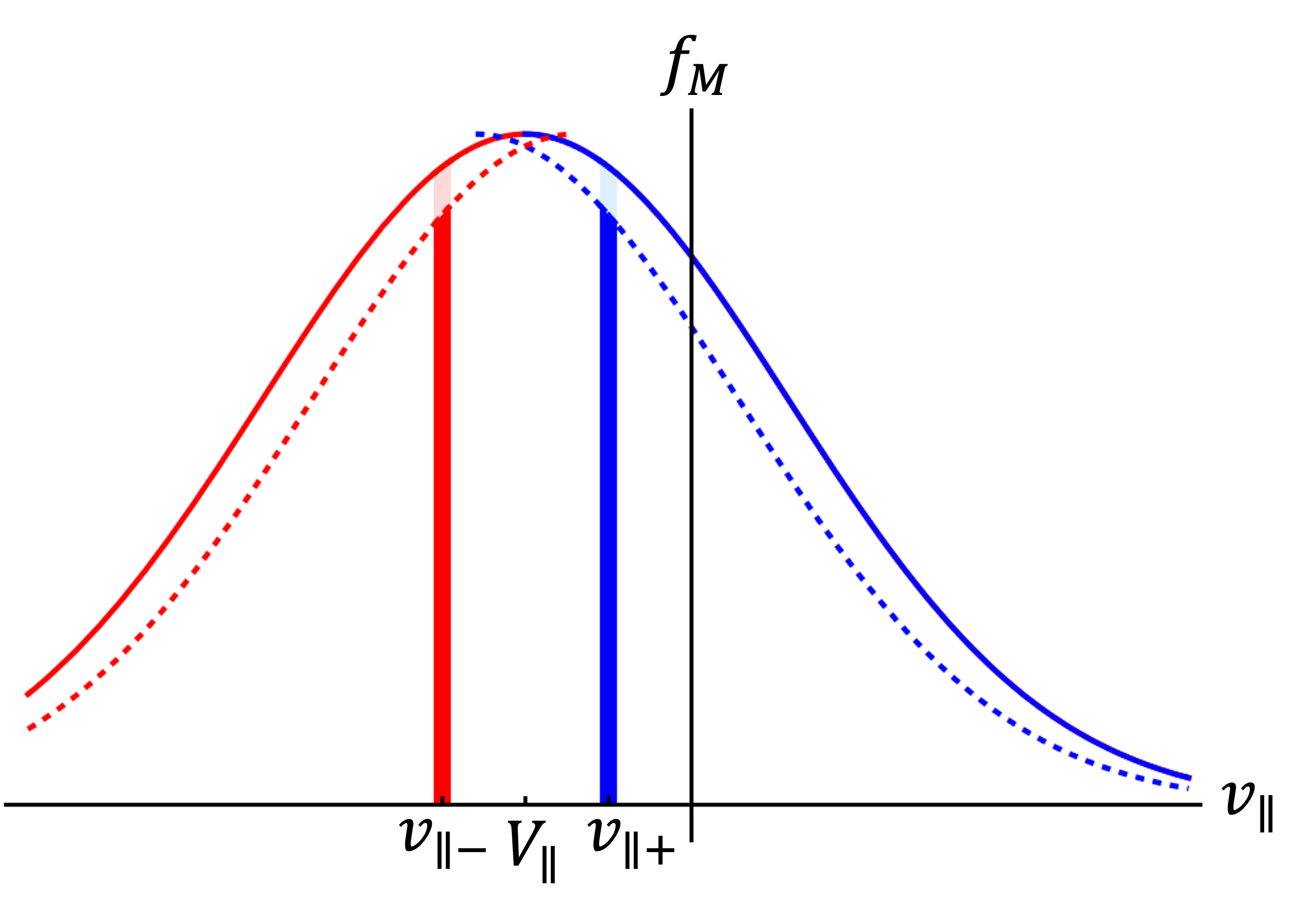}}
    \subfigure[]{ \includegraphics[width=0.48\textwidth]{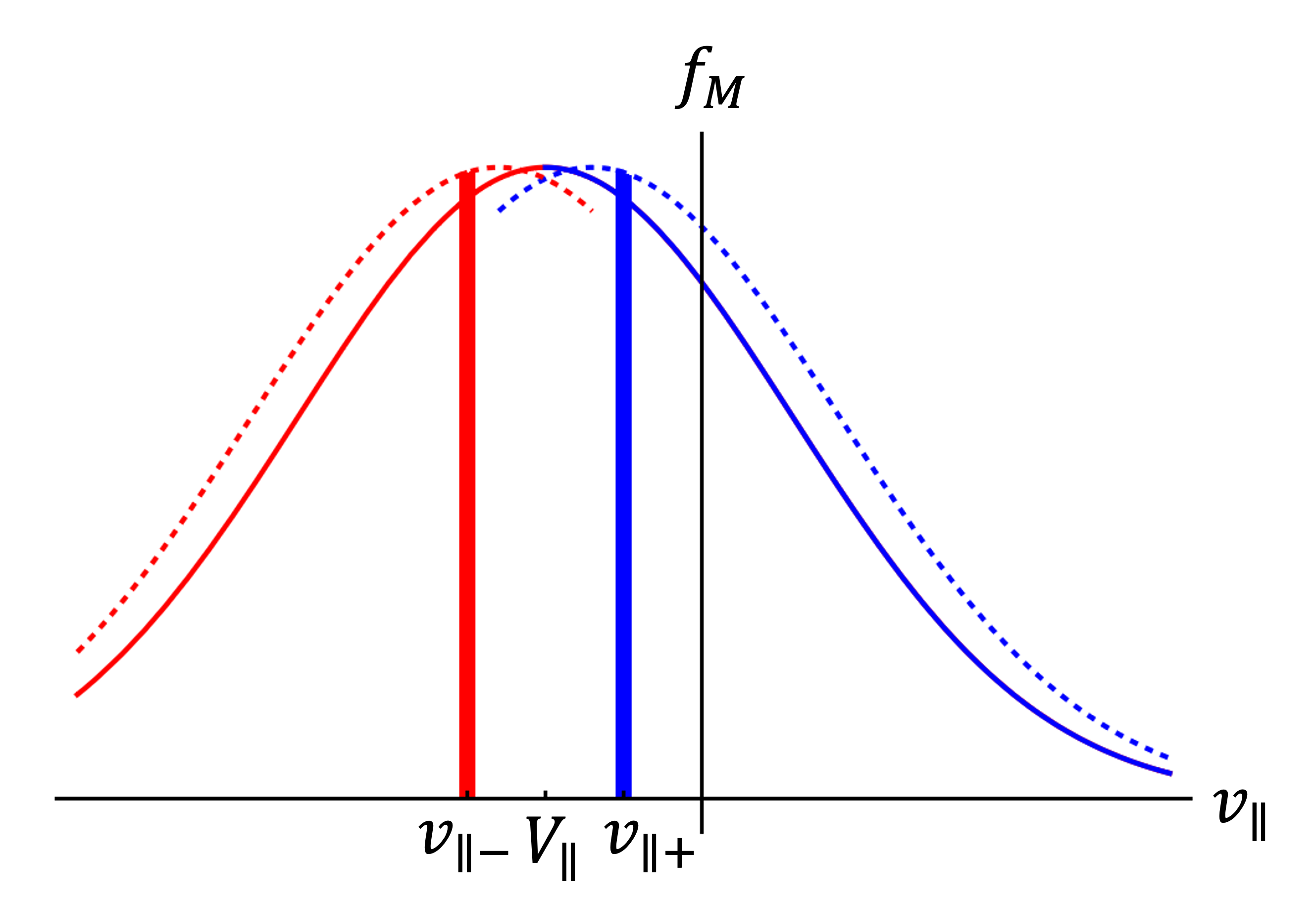}}
     \subfigure[]{ \includegraphics[width=0.35\textwidth]{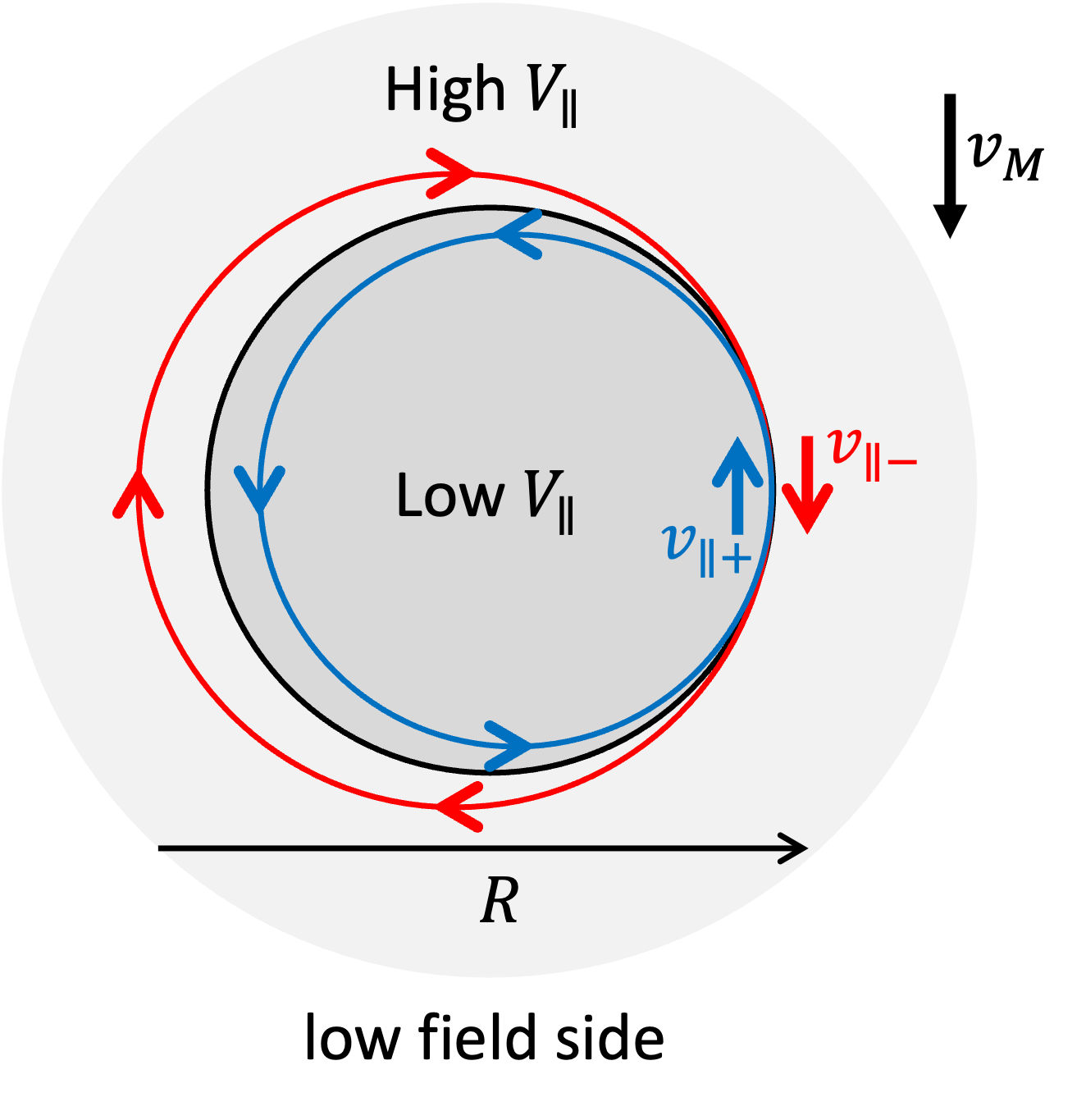}}
    \subfigure[]{ \includegraphics[width=0.32\textwidth]{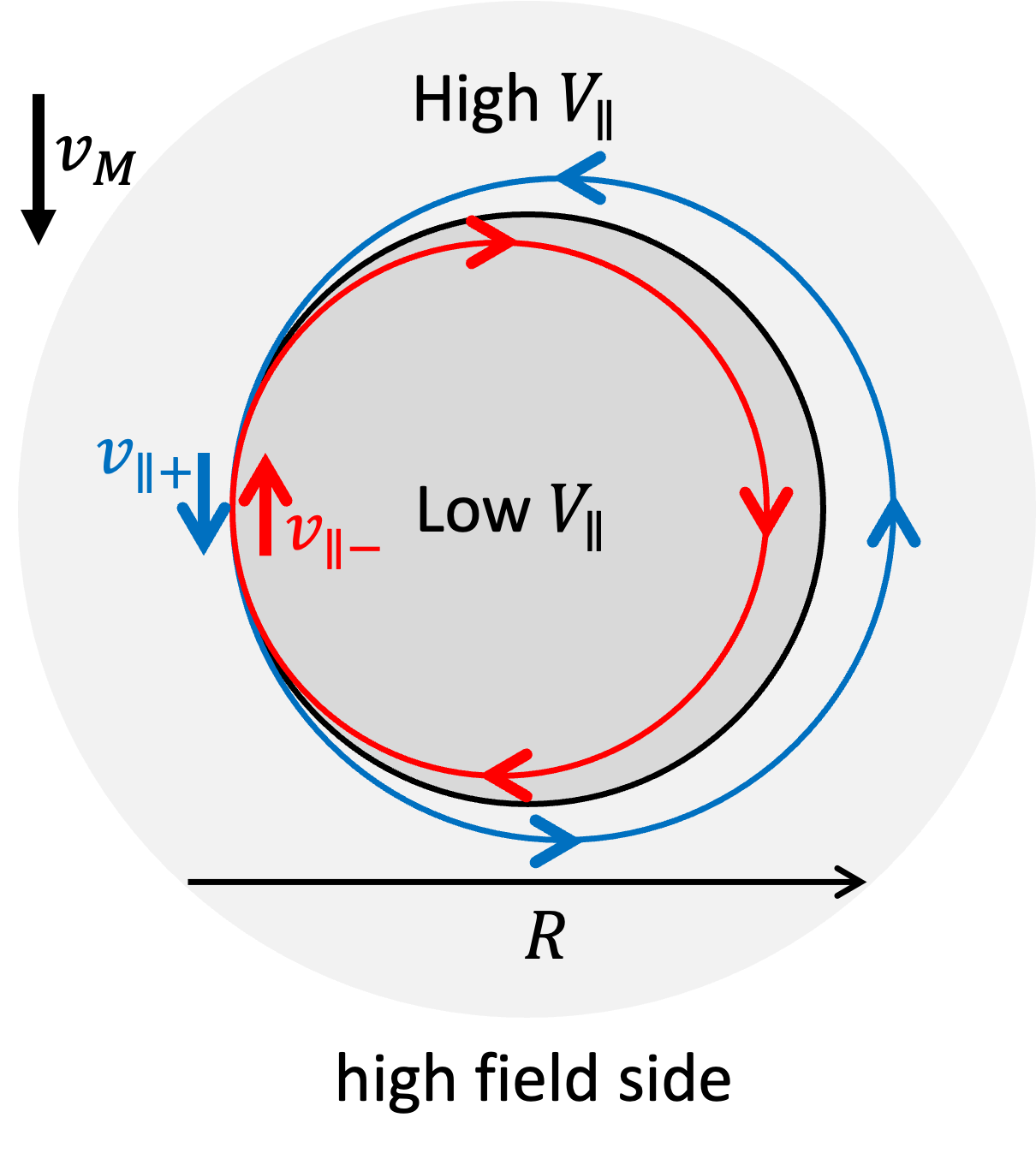}}
    \caption{In this figure, we assume $\partial V_\parallel/\partial \psi>0$. A blue (red) passing particle with parallel velocity $v_{\parallel+}>V_\parallel=-u$ ($v_{\parallel-}<V_\parallel=-u$) on the low field side (a,c) or the high field side (b,d) is circulating in the positive (negative) sense in the poloidal direction. The solid lines represent the Maxwellian on the flux surface of interest. The dashed lines indicate the shifted Maxwellians radially inwards or outwards from the flux surface of interest. On the low field side, blue (red) particles complete their orbits through a region with smaller (larger) mean parallel flow (c), so their average radial position locates them in a region with fewer particles that have a parallel velocity close to $v_{\parallel +}$ ($v_{\parallel -}$) (a). A point of slightly lower density develops on the outboard side. On the high field side, blue (red) particles complete their orbits through a region with larger (smaller) mean parallel flow (d), so their average radial position locates them in a region with more particles that have a parallel velocity close to $v_{\parallel +}$ ($v_{\parallel -}$) (b). A point of slightly higher density develops on the high field side.}
    \label{fig: Vgrad}
\end{figure}

The remaining term is proportional to the mean parallel flow gradient. If the mean parallel flow varies radially, particles with different average radial positions belong to different ion Maxwellian distributions. The difference in mean parallel flow translates into a difference in number of particles. The shift of the Maxwellians between the different flux surfaces are shown in figure \ref{fig: Vgrad} for $V_\parallel=-u$. If, for example, $\partial V_\parallel/\partial \psi$ is positive, the mean parallel flow is smaller on the inside of a flux surface than on the outside. There are fewer particles with positive poloidal velocity $v_{\parallel+}>V_\parallel=-u$ on the low field side (blue particles in figure \ref{fig: Vgrad}(a,c)) because their average radial position is inside the flux surface of interest, where the average flow is smaller than the one in the flux surface of interest. The average radial position of particles with negative poloidal velocity $v_{\parallel -}<V_\parallel=-u$ on the low field side (red particles in figure \ref{fig: Vgrad}(a,c)) locates them in a region of larger mean parallel flow and hence with fewer particles with $v_{\parallel -}$. The low field side develops a region of slightly lower density on the flux surface. On the high field side, the picture reverses. The positively circulating particles with velocities $v_{\parallel +}$ (blue particles in figure \ref{fig: Vgrad}(b,d)) are, on average, in the region where the mean parallel flow is larger. There are more particles with parallel velocities close to $v_{\parallel +}$. The particles with negative poloidal velocity $v_{\parallel -}$ (red particles in figure \ref{fig: Vgrad}(b,d)) belong to a distribution with less particles with $v_{\parallel -}$. The inboard side turns into a region of slightly higher density. Overall, the Boltzmann response of the electrons creates a poloidal potential variation where the outboard side has slightly smaller potential than the inboard side. This effect vanishes in the limit of weak gradients because the mean parallel flow gradient is small.  \par 
 
We discussed the centrifugal force, orbit width asymmetry and flow gradient effects for $V_\parallel=-u$, but they also exist for $V_\parallel\neq -u$. Allowing for $V_\parallel\neq-u$ introduces cross terms that could be attributed to either of the four physical effects: passing particle number asymmetry, centrifugal force, orbit width asymmetry and mean flow gradient. Appendix \ref{Appendix poloidal} has a full list of expressions for the four effects for the purpose of the discussion in section \ref{sec: case study}. For example, we choose to attribute all terms that vanish as $V_\parallel=-u$ to the passing particle number asymmetry.

We now turn our attention to the left hand side of \eqref{phic} and \eqref{phicterms}. The right hand side describes the potential in relation to the magnetic drifts whereas the left hand side is related to the $E\times B$-drift due the poloidal electric field. The origin of the different terms in \eqref{phic} can be traced back to equation (E1) in \cite{Trinczek23}, where the poloidal variation is calculated. An integration is carried out over an expression including the orbit width $\psi_f-\psi$ of particles, which is given in (A6) in \cite{Trinczek23}, and contains terms proportional to $(v_\parallel^2+\mu B)\cos\theta/(v_\parallel+u)$ (the magnetic dirft terms) and terms proportional to $ZeR\phi_1/mr(v_\parallel+u)$ (the $E\times B$-drift terms). 
Since the $E\times B$-drift also contributes to the orbit widths of passing particles, it is equally subject to the passing particle number asymmetry and the mean parallel flow gradient effect. Hence we find contributions to the ion density that are proportional to $\phi_c$ in addition to the usual adiabatic response term, i.e. $en_e/T_e+Z^2n_ie/T_i$, that are also proportional to the gradient of the mean parallel flow or $V_\parallel+u$. The orbit width asymmetry effect is intrinsically a curvature drift effect and unaffected by the poloidal electric field. The centrifugal force is unrelated to drifts and does not appear on the left hand side either. For $V_\parallel+u=0$, the term multiplying $\phi_c$ in \eqref{phicterms} can go negative if $\partial V_\parallel/\partial\psi$ is large enough. When this happens, the $E\times B$-drift induced poloidal variation of the electric potential is strong enough to overcome the effects of the magnetic drifts and the sign of $\phi_c$ reverses. 

\begin{figure}
    \centering
   \subfigure[]{\includegraphics[width=0.48\textwidth]{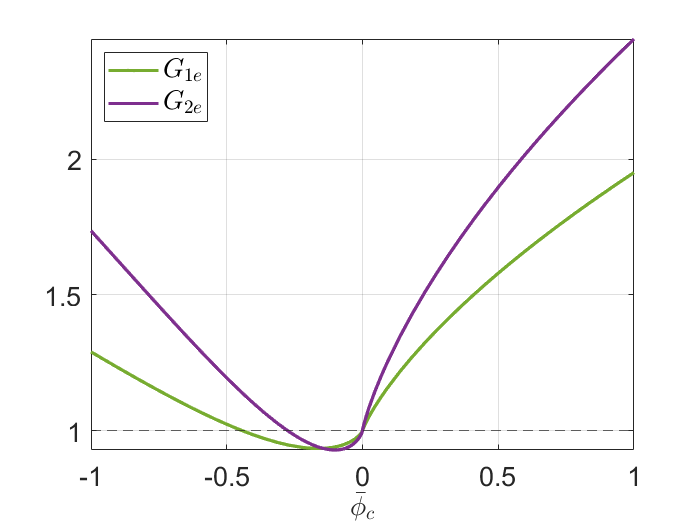}}
   \subfigure[]{ \includegraphics[width=0.48\textwidth]{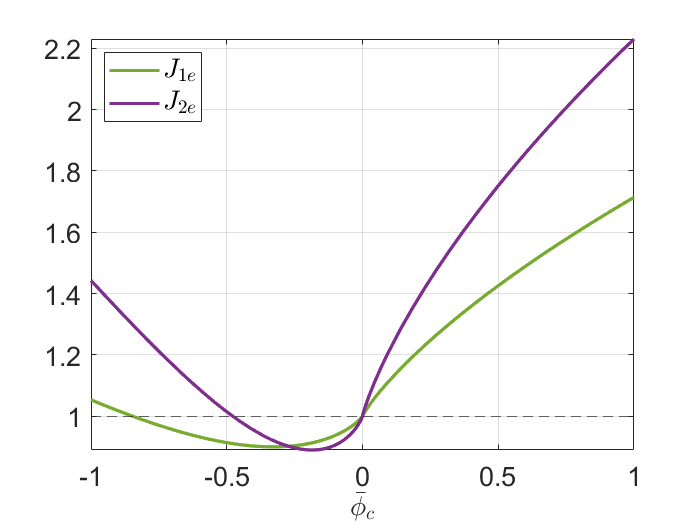}}
       \caption{Modifications $G_{1e}$, $G_{2e}$, $J_{1e}$ and $J_{2e}$ as a function of $\bar{\phi}_c=Ze\phi_cR/T_0r$ as defined in \eqref{G1e}, \eqref{G2e}, \eqref{J1} and \eqref{J2}.}
    \label{fig:Js}
\end{figure}

The poloidal variation of the electric potential enters the transport equations via the four modification functions $J_{1e}$, $J_{2e}$, $G_{1e}$ and $G_{2e}$ which are shown in figure \ref{fig:Js}.
The four functions are all larger than 1 for $\phi_c>0$ in which case the electric potential is slightly higher on the low field side than on the high field side. Consequently, electrons are pushed to the low field side and trapping by the magnetic field is increased. Trapped particles are the main drive of transport, so an increased number of trapped particles gives roughly speaking an enhancement of particle transport and bootstrap current. For a small poloidal variation amplitude, the electrostatic force weakens the magnetic force and less particles are trapped on the low field side. When the poloidal variation becomes negative enough, electrostatic trapping of electrons on the high field side dominates, the number of trapped particles increases again, and the electron transport and bootstrap current are enhanced.\par

\subsection{A case study}\label{sec: case study}
Not only do we know how poloidal variation modifies neoclassical transport properties but we also know where this variation is coming from and how to calculate it self-consistently inside strong gradient regions. To demonstrate this procedure, we first introduce a set of normalised equations and then compare two examples of a pedestal. \par 
The poloidal variation, the neoclassical electron flux and the bootstrap current can be calculated for a given set of profiles for density, temperature, and mean parallel flow. For this purpose, we introduce the normalised variables
\begin{multline}\label{normalised}
    \bar{u}=\sqrt{\frac{m_i}{2T_{i0}}}u, \quad \bar{V}=\sqrt{\frac{m_i}{2T_{i0}}}V_{\parallel },\quad \bar{T}=\frac{T_{i}}{T_{i0}}, \quad \bar{n}=\frac{n_i}{n_{i0}},\\ 
    \bar{\phi}_c=\frac{Ze\phi_{c} R}{T_{i0} r},\quad   \pdv{}{\bar{\psi}}=\frac{I}{\Omega_{i}}\sqrt{\frac{2T_{i0}}{m_i}}\pdv{}{\psi}, \quad \bar{\Gamma}_i=\frac{\Gamma_i}{n_{i0} I\sqrt{\frac{2T_{i0} r}{m_iR}}\frac{\nu_{0}}{\abs{\Omega_{i}}}},\\
     \bar{n}_e=\frac{n_{e}}{Zn_{i0}},\quad  \bar{T}_e=\frac{T_{e}}{T_{i0}}, \quad \bar{j}^B=\frac{\langle j_\parallel^B\rangle_\psi}{Ze n_{i0} \sqrt{\frac{2T_{i0}r}{m_iR}}},\quad \bar{\Gamma}_e=\frac{\Gamma_e}{Zn_{i0} I\sqrt{\frac{2T_{i0} r}{m_iR}}\frac{\nu_{ee,0}}{\abs{\Omega_{e}}}},
\end{multline}
where $n_{i0}$, $T_{i0}$, $\nu_0$, and $\nu_{ee,0}$ are ion density, temperature and ion and electron collision frequencies at a reference flux surface $\bar{\psi}=0$. The electron particle flux in these normalised variables is
\begin{equation}\label{Gammae norm}
    \bar{\Gamma}_e=-1.59 \frac{Z\bar{n}_e^2}{\bar{T}_e^{3/2}}\bigg\lbrace\left[\bar{T}_e \pdv{}{\bar{\psi}}\ln \bar{p}_e-\frac{2}{Z}(\bar{u}+\bar{V})\right]G_{1e}(\bar{\phi}_c,Z)-1.39G_{2e}(\bar{\phi}_c,Z)\pdv{\bar{T}_e}{\bar{\psi}}\bigg\rbrace
\end{equation} 
and the bootstrap current is
\begin{equation}\label{jb norm}
    \bar{j}^B=-1.21\bar{n}_e\bigg\lbrace \left[\bar{T}_e\pdv{}{\bar{\psi}}\ln\bar{p}_e-2\left(\bar{u}+\bar{V}\right)\right]J_{1e}(\bar{\phi}_c,Z)-0.71J_{2e}(\bar{\phi}_c,Z)\pdv{\bar{T}_e}{\bar{\psi}}\bigg\rbrace.
\end{equation}
\par
\begin{figure}
    \centering
    \includegraphics[width=0.8\textwidth]{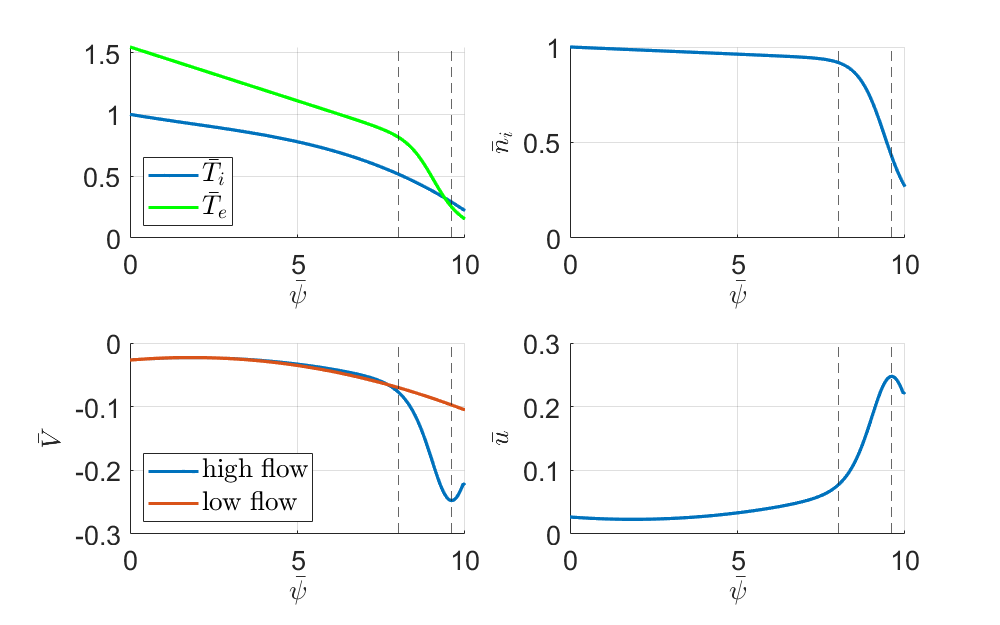}
    \caption{Input profiles for density, temperature, mean parallel flow and radial electric field.}
    \label{fig:Profiles}
\end{figure}

The point $\bar{\psi}=0$ is not the magnetic axis but a point where gradients are sufficiently small that usual neoclassical theory can be used. The radial electric field in the pedestal is often assumed to be mostly determined by the pressure gradient \citep{Kagan08,McDermott09,viezzer2013}. For this example, we assume
\begin{equation}\label{force balance}
    Zen_i\pdv{\Phi}{\psi}+\pdv{p_i}{\psi}=0,
\end{equation}
from which $u$ can be calculated as shown in figure \ref{fig:Profiles}.\par 
The mean parallel flow no longer follows from the neoclassical ion particle flux equation in strong gradient regions because for strong radial electric fields the mechanism of flow damping is no longer dominant \citep{Trinczek23}. The parallel flow has to be determined via a balance between flow damping and momentum transport. We discuss the intricacy of the calculation of the mean parallel flow in the next section. In this section, we compare two different, sensible cases for the mean parallel flow. In the first case,
\begin{equation}
    V_\parallel=-\frac{IT_i}{m_i\Omega_i}\left(\pdv{}{\psi} \ln p_i +\frac{Ze}{T_i}\pdv{\Phi}{\psi}-1.17 \pdv{}{\psi}\ln T_i\right).
\end{equation}
We call this case "low flow" because this expression is the typical result for the mean parallel flow in the low flow regime of weak gradient neoclassical theory. We call the second case "high flow" because of the choice
\begin{equation}
    V_\parallel=-u,
\end{equation}
which is the result for the mean parallel flow in the high flow regime of weak gradient neoclassical theory. Thus, the system is not studied in the usual low flow or high flow limits when comparing the two example cases -- rather we choose these two $V_\parallel$ profiles as reasonable assumptions for the ion mean parallel flow. \par

\begin{figure}
    \centering
    \subfigure[]{\includegraphics[width=0.48\textwidth]{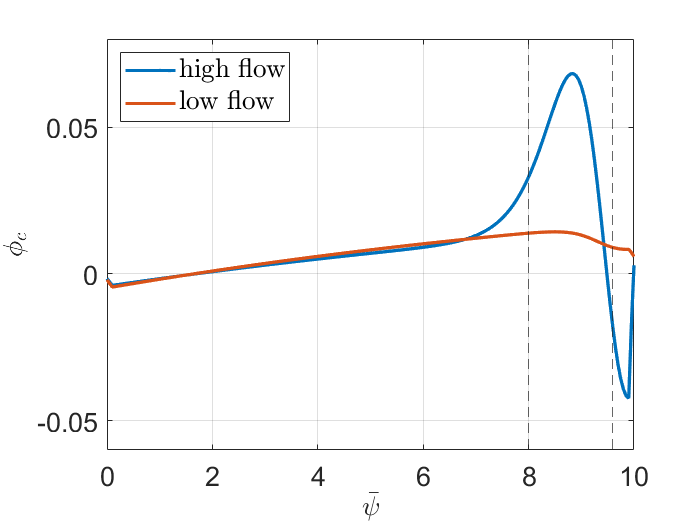}}
 \subfigure[]{\includegraphics[width=0.48\textwidth]{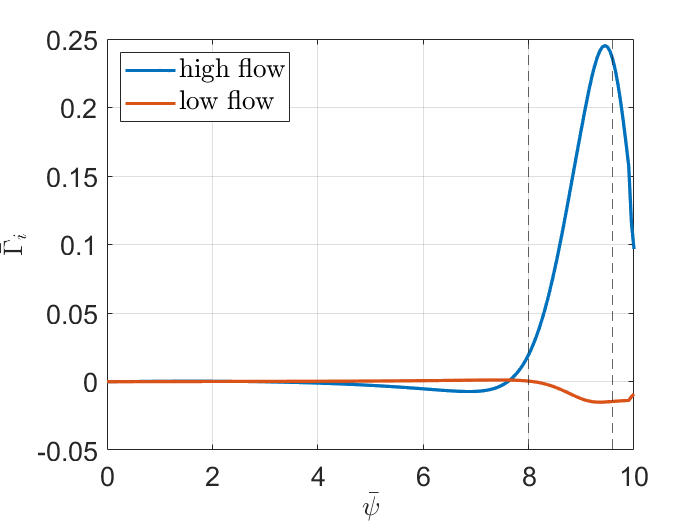}}
    \caption{Amplitude of the poloidal variation of the electric potential and neoclassical ion particle flux for the example profiles in figure \ref{fig:Profiles}.}
    \label{fig:Gamma}
\end{figure}

\begin{figure}
    \centering
     \subfigure[]{\includegraphics[width=0.48 \textwidth]{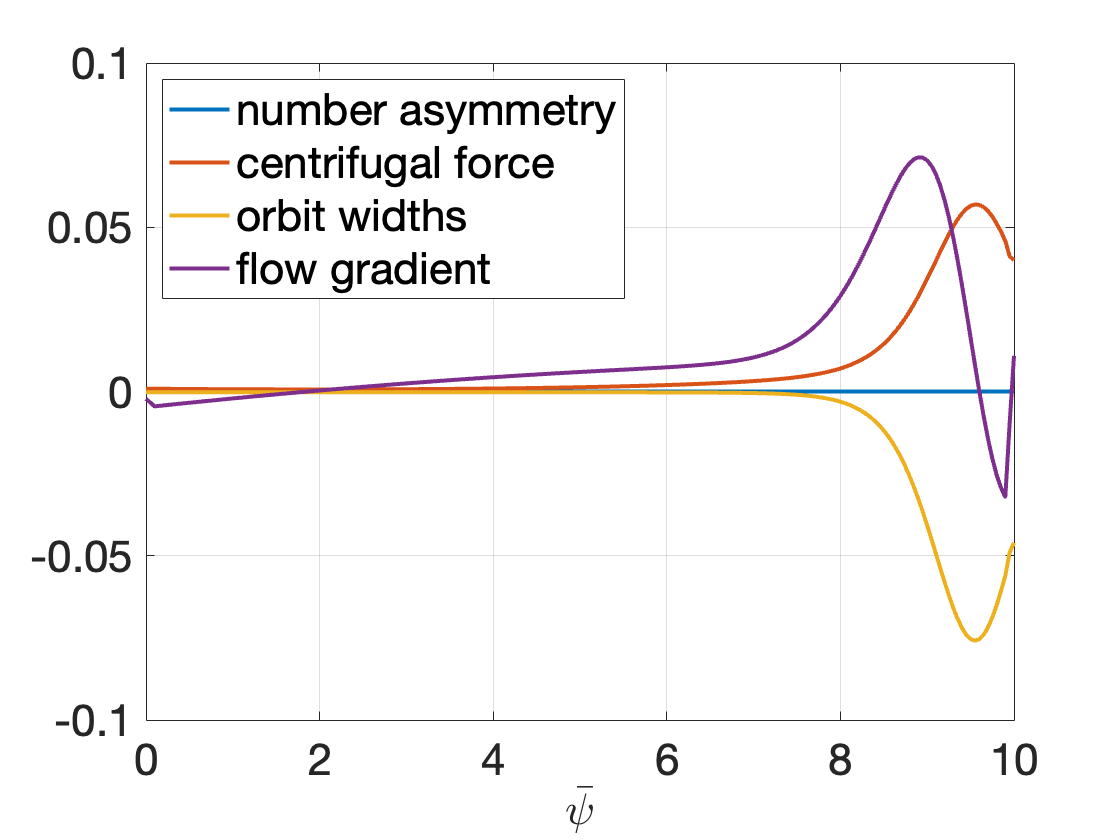}}
     \subfigure[]{\includegraphics[width=0.48 \textwidth]{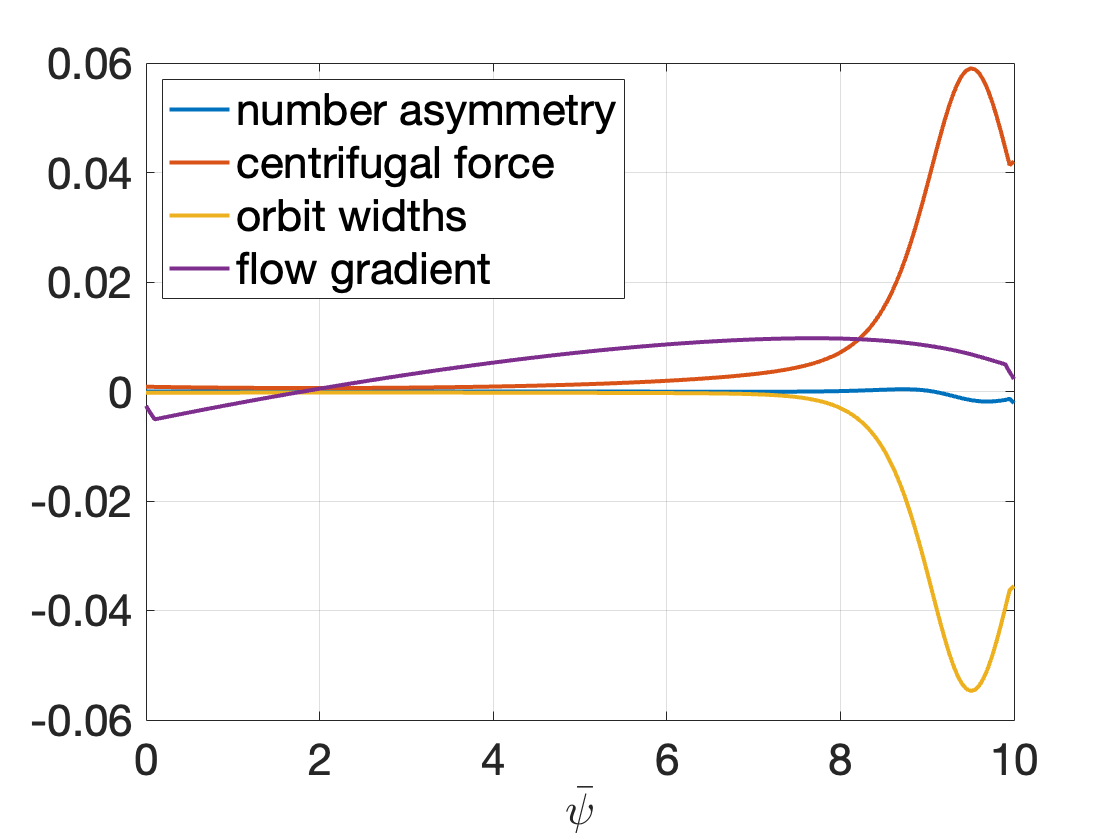}}
    
    \caption{The poloidal potential variation amplitude can be split up into four different contributions, associated with the effect of passing particle number asymmetry, centrifugal force, mean parallel flow gradient, and orbit width asymmetry. The mathematical expressions we used for this figure are summarized in Appendix \ref{Appendix poloidal}. The blue line shows the normalised contribution from the passing particle number asymmetry, the red line shows the piece due to the centrifugal force, the yellow line shows the normalised contribution from the asymmetry in the orbit width and the purple line shows the normalised contribution from the mean parallel flow gradient. Figure (a) shows the individual contributions in the "high flow" example and figure (b) shows the individual contributions in the "low flow" example.}
    \label{fig: phicterms}
\end{figure}

The example profiles for temperatures and density for $Z=1$ displayed in figure \ref{fig:Profiles} are taken from \cite{Viezzer17}. The analytical formulas of the profiles are given in Appendix \ref{sec: Input formulas}. The poloidal variation of the electric potential as well as ion transport profiles were calculated by \cite{Trinczek23} for the set of input profiles in figure \ref{fig:Profiles}. We replicate the results for the neoclassical ion particle flux and $\bar{\phi}_c$ in figure \ref{fig:Gamma}. For weak gradients, $\bar{\phi}_c$ is very small, as expected. The different contributions to the potential amplitude as discussed in section \ref{sec: phic effects} are plotted individually for the "high flow" and "low flow" case in figure \ref{fig: phicterms}. The effect of passing particle number asymmetry vanishes exactly in the "high flow" case because $V_\parallel+u=0=J$ and gives only a small contribution in the "low flow" case. Interestingly, centrifugal force and orbit width asymmetry effects balance each other such that the main contribution to the total poloidally varying part of the potential is derived from the mean parallel flow gradient effect in both cases. The cancellation between centrifugal force and orbit width asymmetry is partially due to our choice of force balance in \eqref{force balance}.


\begin{figure}
    \centering
    \subfigure[]{\includegraphics[width=0.48\textwidth]{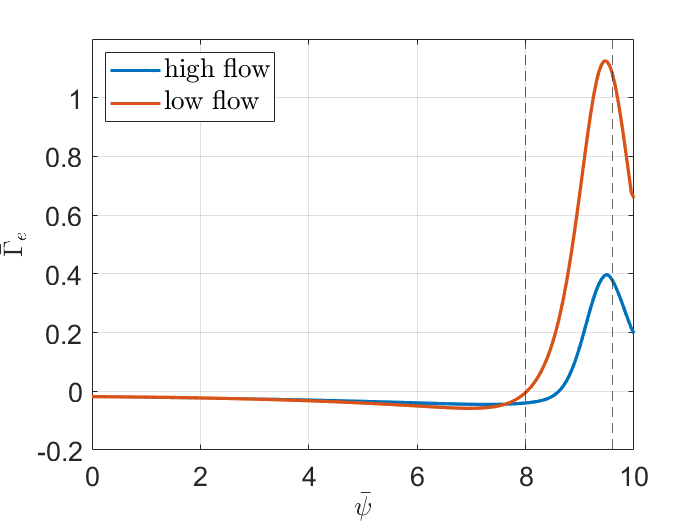}}
 \subfigure[]{\includegraphics[width=0.48\textwidth]{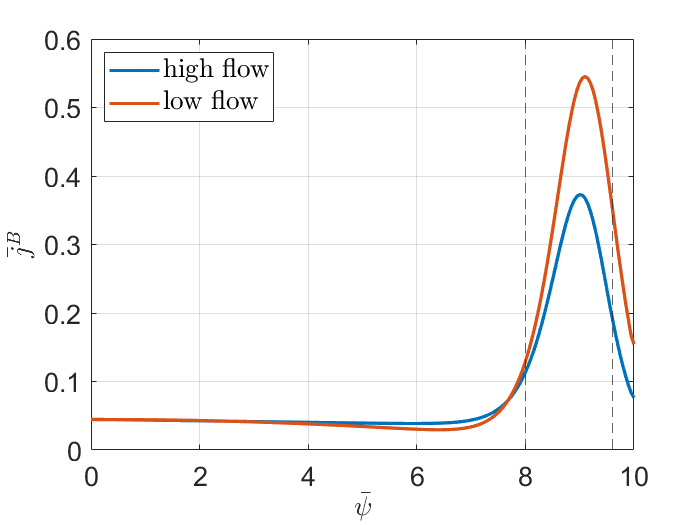}}
    \caption{Neoclassical electron flux and bootstrap current for the example profiles in figure \ref{fig:Profiles}.}
    \label{fig:Gammae}
\end{figure}

Using the set of input profiles in figure \ref{fig:Profiles}, we find the profile of the neoclassical electron flux and bootstrap current as shown in figure \ref{fig:Gammae}.

The overall particle fluxes of ions and electrons have to balance each other to satisfy ambipolarity. The total fluxes consist of a turbulent and a neoclassical contribution. \cite{Trinczek23} showed that, due to the strong gradients considered, the neoclassical ion and electron particle fluxes need not balance each other. That is why the "high flow" ion flux can be large as shown in figure \ref{fig:Gamma}(b). Note that $\Gamma_i$ and $\Gamma_e$ are normalized differently in \eqref{normalised} when comparing the neoclassical particle fluxes for ions and electrons. This is different from weak gradient neoclassical theory, where the neoclassical ion and electron fluxes have to be equal and the lowest order ion particle flux has to vanish. As pointed out in \cite{Trinczek23}, a non-zero lowest order neoclassical ion transport in a strong gradient system requires a source of parallel momentum that could be provided via interactions with turbulence. \par
The neoclassical electron particle flux grows significantly in the strong gradient region that is bracketed between the dashed lines in both the "low flow" case and the "high flow" case. Interestingly, the neoclassical particle flux of electrons in the "high flow" case is smaller than in the "low flow" case as opposed to the neoclassical ion flow, where the picture was reversed and the larger particle flux was found in the "high flow" case. Similarly, the bootstrap current grows significantly in the strong gradient region and is smaller in the "high flow" case than in the "low flow" case. The difference between the "high flow" and the "low flow" case can be traced back to the difference in the coefficients $G_{1e}$, $G_{2e}$, $J_{1e}$, and $J_{2e}$. For positive $\bar{\phi}_c$, $G_{2e}>G_{1e}$ and $J_{2e}>J_{1e}$. From equations \eqref{Gammae norm} and \eqref{jb norm}, one can see that the term proportional to $G_{2e}$ ($J_{2e}$) decreases the neoclassical electron flux (bootstrap current), whereas the terms proportional to $G_{1e}$ ($J_{1e}$) increase it. The poloidal variation of the potential is stronger in the "high flow" case and thus the difference between $G_{1e}$ ($J_{1e}$) and $G_{2e}$ ($J_{2e}$) is larger. The modifications due to large gradients reduce the electron particle flux (bootstrap current) in the "high flow" case more strongly than in the "low flow" case. Additionally, the term multiplying $G_{1e}$ ($J_{1e}$) is smaller in the "high flow" case because $\bar{V}+\bar{u}$ vanishes exactly. 
\par
A comparison of the "low flow" and "high flow" cases with the respective results in the weak gradient limit \eqref{Gammae wg} and \eqref{jb wg} shows the significance of the poloidal variation modification. All four comparisons are shown in figure \ref{fig:compare}. The equations for the weak gradient limit that we use are those in Appendix \ref{sec:weak comp}. In the weak gradient high flow limit, the poloidal variation reduces to the contribution from centrifugal forces \eqref{phi centri}.
\begin{figure}
    \centering
    \subfigure[]{\includegraphics[width=0.45\textwidth]{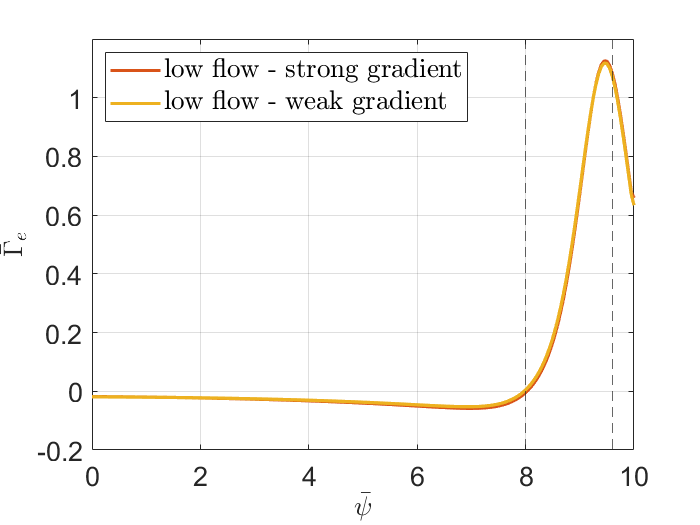}}
    \subfigure[]{\includegraphics[width=0.45\textwidth]{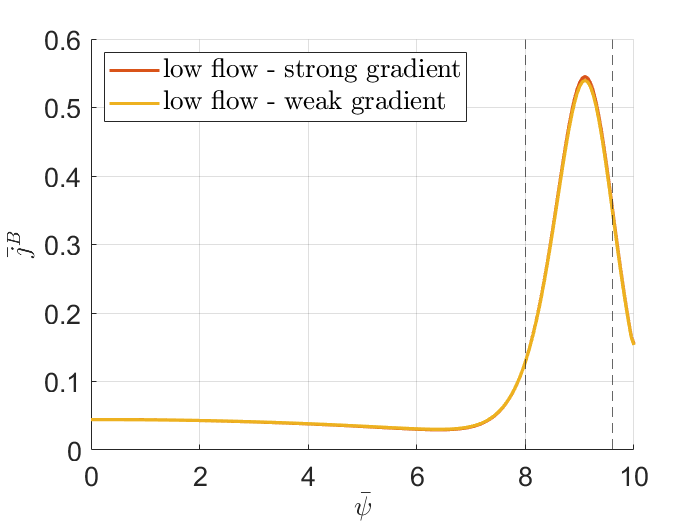}}
    \subfigure[]{\includegraphics[width=0.45\textwidth]{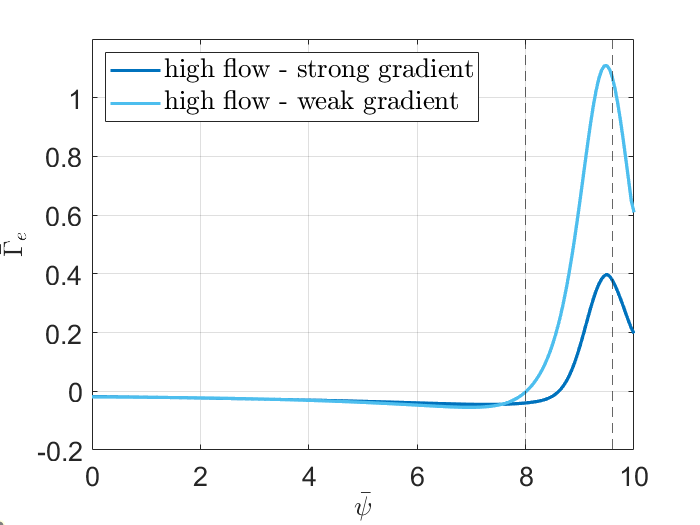}}
    \subfigure[]{\includegraphics[width=0.45\textwidth]{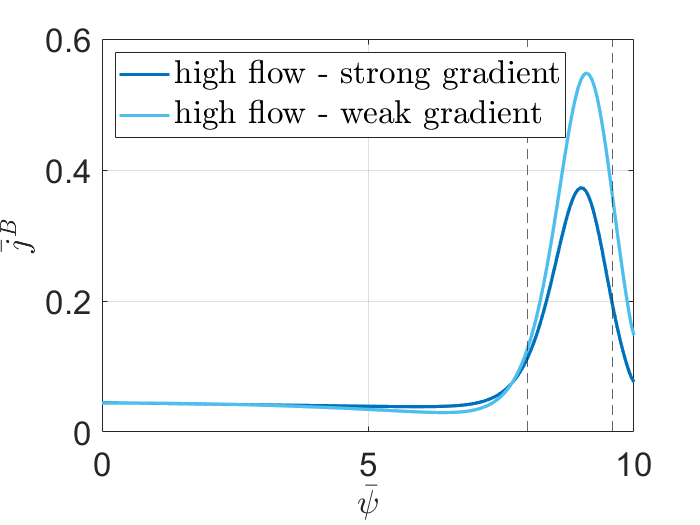}}
    \caption{Comparison of strong gradient and weak gradient neoclassical electron particle fluxes and bootstrap current for "high flow" and "low flow".}
    \label{fig:compare}
\end{figure}
In the "low flow" case, the differences between strong gradient and weak gradient neoclassical theory are small. In the "high flow" case, the difference between weak gradient and strong gradient theory are significant. The respective maxima of the particle flux and bootstrap current are reduced by a factor of $\bar{\Gamma}_e/\bar{\Gamma}_e^{wg,hf}\simeq0.36$ and $\bar{j}^B/\bar{j}^{B,wg,hf}\simeq 0.68$. \par 

These results are not universal and they are highly dependent on $V_\parallel$ (see section \ref{Mean parallel flow sec}). In fact, for $T_e=T_i$, using the $T_e$ and $n_i$ profiles in figure \ref{fig:Profiles}, the "low flow" case gives an increase in bootstrap current of the order of 5\%. It is thus possible to construct cases that predict a higher or lower bootstrap current in comparison to weak gradient neoclassical theory. Less current drive might be required if the bootstrap current in the pedestal is in fact larger than assumed. However, a larger bootstrap current might also lead to more instabilities and is not necessarily favorable. 
\par
It is clear from figure \ref{fig:Gammae} that the strong gradient modifications are very different in the "high flow" and "low flow" examples. The strong dependence on the mean parallel flow profile is reflected in the amplitude of the poloidal variation of the electric potential in figure \ref{fig:Gamma}. The mean parallel flow is not only relevant for the enhancement or reduction of fluxes and bootstrap current but it also leads to qualitative differences as can be seen from the sign change of $\phi_c$ in the "high flow" case (see figure \ref{fig:Gamma}). More work is needed to accurately determine the mean parallel flow and its impact on neoclassical transport in strong gradient regions.

\section{Mean parallel flow} \label{Mean parallel flow sec}
In the preceding section, we chose profiles for density and temperature, assumed radial force balance between the radial electric field and the pressure gradient, and compared the fluxes and bootstrap current for two different example profiles for the mean parallel flow. This procedure begs the question if one can reverse the process and determine the mean parallel flow, density and temperature for a given set of particle, momentum and heat sources. The calculation of the mean parallel flow is particularly interesting as experimental profiles for the mean parallel flow of bulk ions are difficult to obtain. Furthermore, the strong gradient effects presented here depend strongly on the mean parallel flow profile, as demonstrated by figure \ref{fig:compare}. Without a full understanding of the mean parallel flow, it is unclear if strong gradient effects modify weak gradient neoclassical theory significantly ("high flow" case in figure \ref{fig:compare}) or not ("low flow" case in figure \ref{fig:compare}). It turns out that solving for the mean parallel flow is not straightforward.\par
\cite{Trinczek23} derived the parallel momentum equation
\begin{equation}\label{parallel momentum}
    \pdv{}{\psi}\left(m_i u\Gamma_i\right)+\frac{m_i\Omega_i}{I}\Gamma_i=-\gamma,
\end{equation}
where $\gamma$ is the source of parallel momentum. \par
First, we consider the case of a purely neoclassical pedestal. Without turbulent transport, ion and electron neoclassical particle flux have to be ambipolar and thus the neoclassical ion particle flux to this order has to vanish. The left side of \eqref{parallel momentum} vanishes identically and no information can be extracted from this equation. One has to go to higher order in the parallel momentum equation derivation. This higher order equation could then be used to determine the mean parallel flow. This work is ongoing and will be presented in the future.\par
If we allow for a non-vanishing ion neoclassical particle flux to this order, i.e. a strong turbulent electron particle flux exists to provide ambipolarity, equation \eqref{parallel momentum} is not identically zero. However, the mean parallel flow does not appear in this equation. For a given parallel momentum source and particle source, which sets $\Gamma_i$, one could determine $u$, but not $V_\parallel$. Instead, the particle flux equation needs to be solved to determine the mean parallel flow, where the ion neoclassical particle flux is given in \eqref{Gamma}. If we solve this equation for a given density and temperature profile, a known particle flux $\Gamma_i$ and a profile for $u$ as determined from the parallel momentum equation \eqref{parallel momentum}, one should in principle be able to determine the mean parallel flow. Indeed, one can try to integrate the profile of the mean parallel flow profile for a given set of profiles of $T_i$, $T_e$, $n_i$, $u$, and a boundary value for $V_\parallel$ by solving \eqref{Gamma} for $\partial V_\parallel/\partial \psi$ at every radial point. However, this procedure does not always yield a solution, that is, solutions do not exist for all boundary conditions and sources. On the right hand side of \eqref{Gamma}, there is an explicit linear dependence on the gradient of the mean parallel flow. However, in addition to this dependence, the functions $G_1$ and $G_2$ depend on the gradient of the mean parallel flow through the poloidally varying part of the electric potential $\phi_c$ which is given in \eqref{phic}. Through the coupling between the particle flux equation, equation \eqref{Gamma}, and the poloidal variation of the potential in the argument of $G_1$ and $G_2$, \eqref{phic}, the equation is highly nonlinear in the gradient of the mean parallel flow. \par
\begin{figure}
    \centering
    \includegraphics[width=0.7\textwidth]{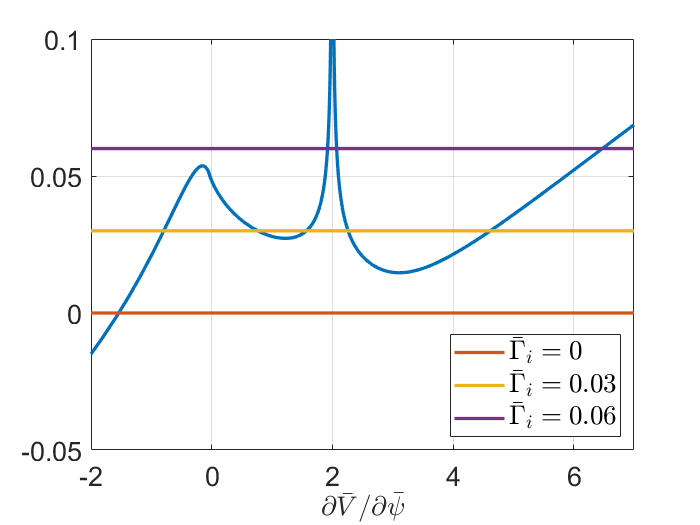}
    \caption{We plot in blue the right hand side of \eqref{Gamma} as a function of $\partial \bar{V}/\partial \bar{\psi}$ for $\bar{T}=\bar{T}_e=0.47$, $\bar{n}=0.87$, $\bar{u}=0.1$, $V=-0.1$, $\partial \bar{T}/\partial \bar{\psi}=-0.13$, $\partial \bar{n}/\partial \bar{\psi}=-0.14$, $\partial \bar{u}/\partial\bar{\psi}=0.09$. Solutions to \eqref{Gamma} exist for specific values of the neoclassical ion particle flux but the number of roots change with the value of $\Gamma_i$. Solutions disappear or run away when $\Gamma_i$ changes. Five solutions exist for $\bar{\Gamma}_i=0.03$ but only three solutions exist for $\bar{\Gamma}_i=0.06$ and one for $\bar{\Gamma}_i=0$.}
    \label{fig: VFail}
\end{figure}
In figure \ref{fig: VFail} we plot the right hand side of \eqref{Gamma} in blue as a function of $\partial V_\parallel/\partial \psi$ and compare it to different values of $\Gamma_i$, keeping everything else fixed. The blue curve has an asymptote where $\phi_c$ goes to infinity because the factor multiplying $\phi_c$ on the left hand side of \eqref{phic} vanishes. Up to five roots can be found for a given value of the particle flux. Due to radial dependence of $\Gamma_i$, taking a step in radius can be thought of as moving from one horizontal line to another, although, in our examples, temperature, density, and mean flow are not constant and the blue curve itself would change its shape when taking a step in radius. However, if $\Gamma_i$ changes from, for example, $\bar{\Gamma}_i=0.03$ to $\bar{\Gamma}_i=0.06$, the number of roots reduces from 5 to 3. The number of solutions reduces to one when $\Gamma_i$ changes from $\bar{\Gamma}_i=0.03$ to $\bar{\Gamma}_i=0$. Solutions seem to disappear as $\Gamma_i$ changes, holding everything else fixed. Some solutions to \eqref{Gamma} run off to infinity and give unphysical solutions such as the rightmost solution in the example in figure \ref{fig: VFail}, for which increasing $\bar{\Gamma}_i$ from $0.03$ to $0.06$ leads to an increase in $\partial \bar{V}/\partial\bar{\psi}$, or the leftmost solution, for which decreasing $\bar{\Gamma}_i$ from 0.03 to 0 gives a decrease in $\partial \bar{V}/\partial\bar{\psi}$. This example shows how for zero neoclassical ion particle flux, only a solution with very strong mean parallel flow gradient and consequently strong rotation exists for our choice of density, temperatures and radial electric field. The sources need to be constructed such that a sensible solution exists at each radial point for a changing set of parameters. It is difficult to find the correct boundary conditions and source terms to construct sensible, non-singular solutions. \par
Solutions that extend all across the pedestal only exist for specific boundary conditions and source terms. For this reason, we limit ourselves to studying realistic example profiles for $V_\parallel$ in this article. We will investigate the derivation of $V_\parallel$ from higher order parallel momentum conservation in the context of purely neoclassical transport further in future work.

\section{Radial electric field}\label{Radial electric field}
Previously, the assumption of radial force balance in \eqref{force balance} was used to determine the radial electric field from the pressure profile. Equation \eqref{force balance} is an assumption based on experimental observations that the radial electric field is mostly set by the pressure gradient \citep{Kagan08, McDermott09, viezzer2013}. Despite the experimental motivation behind this assumption, enforcing radial force balance does not enable us to calculate the mean parallel flow and close the system of equations. However, the assumption of radial force balance can be dropped in a system where the neoclassical ion particle flux $\Gamma_i$ is small and balances the neoclassical electron particle flux $\Gamma_e$. We show that in such a case the radial electric field can be determined for a given profile of the mean parallel flow. \par 

In the absence of turbulence and external injection, no source of parallel momentum is present. Therefore, the right hand side of the parallel momentum equation \eqref{parallel momentum} can be set to zero. Equation \eqref{parallel momentum} then predicts an exponential decay of the neoclassical ion particle flux in the pedestal \citep{Trinczek23}. This is consistent with neoclassical ambipolarity where the neoclassical fluxes are equal in the absence of turbulence, forcing $\Gamma_i\simeq0$. In such a turbulence--free pedestal, the left hand side of \eqref{Gamma} vanishes to lowest order and the right hand side of \eqref{Gamma} can be solved for $u$, i.e. the radial electric field for a given set of density, temperature, and mean parallel flow profiles. Although this does not solve the problem of determining the mean parallel flow, this procedure does not rely on the assumption of radial force balance. Instead, the argument for neoclassical particle flux balance $\Gamma_i\sim \Gamma_e$ follows from the lack of turbulent fluxes. The balance does not follow automatically from the neoclassical equations like in the weak gradient regions.\par

\begin{figure}
    \centering
    \includegraphics[width=0.9\linewidth]{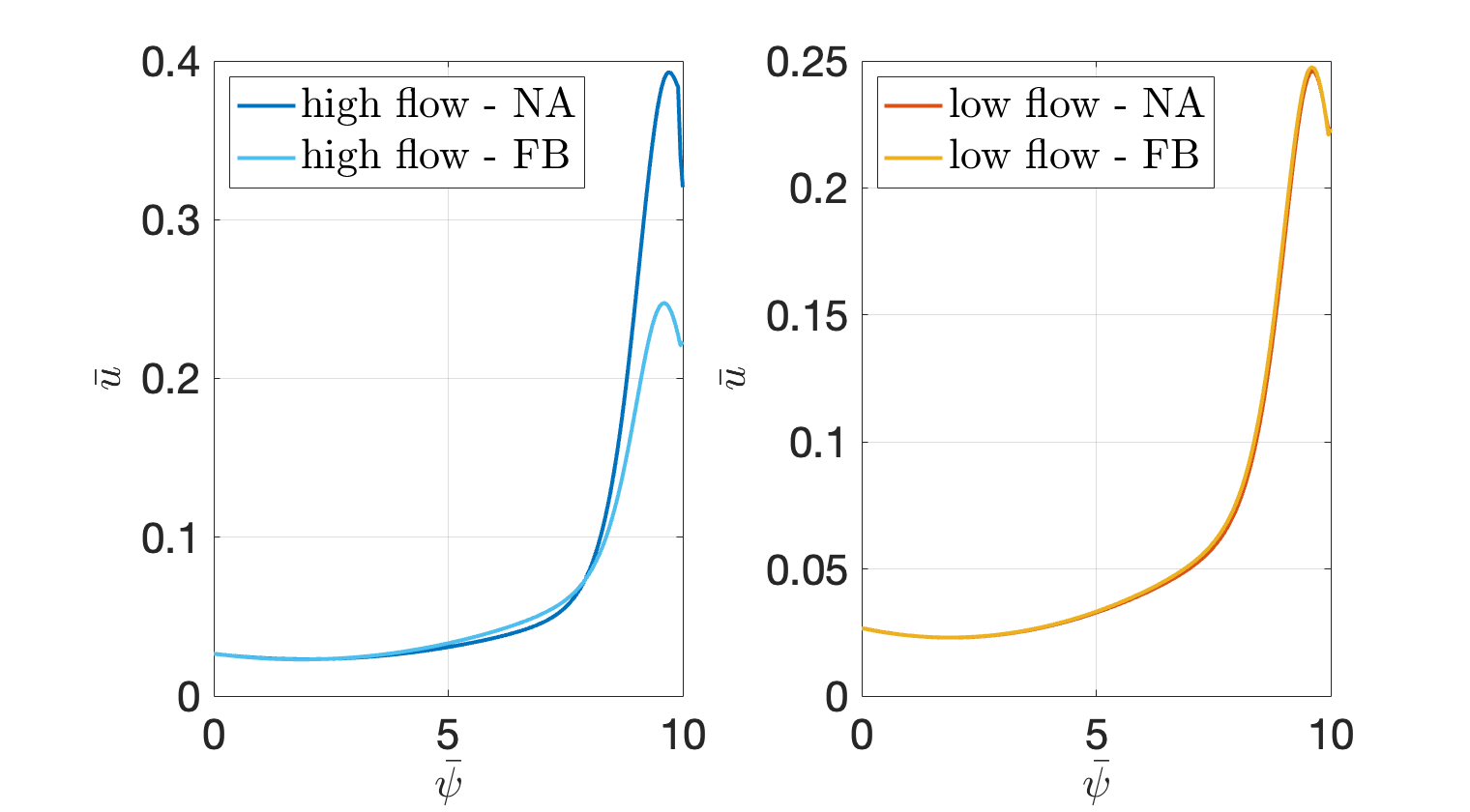}
    \caption{We compare the profile of $\bar{u}$ for the case where the radial electric field is determined by neoclassical ambipolarity (NA), so $\Gamma_i=0$, to the previous approach using radial force balance (FB) in \eqref{force balance} for the "high flow" and "low flow" example.  }
    \label{fig: SolveForEr}
\end{figure}

Turning back to the example profiles for density and temperature in figure \ref{fig:Profiles}, we can determine the radial electric field. We set the left hand side in \eqref{Gamma} to zero as predicted by neoclassical ambipolarity and solve for $u$ for the "low flow" and the "high flow" mean parallel flow example, respectively. The results for $u$ are shown in figure \ref{fig: SolveForEr}. We compare the solution using neoclassical ambipolarity to $u$ as predicted by force balance \eqref{force balance}.
In the "low flow" case, the radial electric field as determined by neoclassical ambipolarity and radial force balance agree very well in a purely neoclassical pedestal. In the "high flow" case, the radial electric field as determined by neoclassical ambipolarity exceeds the radial electric field that follows from \eqref{force balance} by a factor of $\bar{u}^{hf,NA}/\bar{u}^{hf,FB}\simeq 1.59$ at the maximum values. The radial electric field based on neoclassical ambipolarity is larger in a turbulence--free pedestal than one expects from radial force balance. Again, the choice of the mean parallel flow is crucial. For the purely neoclassical pedestal, \eqref{parallel momentum} vanishes exactly and a higher order calculation is needed to determine the mean parallel flow and close the system of equations self-consistently.

\section{Conclusion}\label{sec:conclusion}
Strong gradient effects cause poloidal variation of the electric potential and change the mean flow according to \cite{Trinczek23}. Both of these effects modify the neoclassical transport of electrons and the bootstrap current. In this paper, these modifications were derived, explained and studied using example profiles. \par 
Consecutive expansions in small collisionality, mass ratio and large aspect ratio facilitate an analytical treatment of electron physics in strong gradient regions such as the pedestal and internal transport barriers. Expressions for the electron distribution function were derived using fixed-$\theta$ variables and a jump condition approach. The resulting neoclassical electron particle flux equation is different from the one given by weak gradient neoclassical theory due to poloidal variation in the electric potential, and differences in the mean parallel flow caused by strong gradient effects. The bootstrap current can be derived using self-adjointness of the collision operator or alternatively using the same jump condition approach (as in Appendix \ref{Appendix bootstrap}). The bootstrap current experiences modifications driven by the poloidal variation of the electric potential and the changes in the parallel flow just like the electron particle transport. \par
The poloidal variation of the electric potential was revisited and studied in more detail. The poloidal variation originates from four different strong gradient effects. The four effects are the asymmetry in the number of passing particles, the centrifugal force, the asymmetry in the orbit width and the gradient of the mean parallel flow. We have provided physical pictures for all four of them. \par 
The neoclassical electron particle flux and the bootstrap current can be calculated for a given set of density, temperature and mean flow profiles. Assuming radial force balance between the pressure gradient and the electric field, "low flow" and "high flow" profiles were studied and compared to weak gradient neoclassical predictions. The relevance of the four strong gradient effects was evaluated. The passing particle number asymmetry effect is relatively small in the "high flow" and the "low flow" case. Centrifugal and orbit width effects have significant contributions that balance each other. The total poloidal variation is mostly set by the mean parallel flow gradient effect. The "low flow" electron flux and bootstrap current are larger than in the "high flow case", which is related to the sign of $\phi_c$ and the size of $V_\parallel+u$. No significant changes from weak gradient neoclassical theory were observed in the "low flow" example. The "high flow" weak gradient solutions for the bootstrap current overestimate the bootstrap current in the pedestal by roughly a factor of two and the neoclassical electron particle flux by a factor of three. For a different choice of temperature profiles, an increased bootstrap current in the "low flow" example can be observed, a result that demonstrates strong gradient effects can in principle cause an increased or decreased prediction of the bootstrap current. \par 
We showed that different mean parallel flow profiles can lead to very different outcomes for electron transport and bootstrap current. In general, solutions for the parallel flow only exist for specific sources and boundary conditions. Momentum conservation to higher order is required to determine the mean parallel flow for the purely neoclassical pedestal. More work is required to understand the mean parallel flow in the pedestal. \par
In the case of a purely neoclassical pedestal, the ion neoclassical particle flux vanishes to lowest order. For a known parallel flow profile, neoclassical ambipolarity can be used to determine the radial electric field in this turbulence--free case. 
\section*{Acknowledgments}

\section*{Funding}
This work was supported by the U.S. Department of Energy (S.T. and F.I.P., contract number DE-AC02-09CH11466 and P.J.C., contract number DE-FG02-91ER-54109). The United States Government retains a non-exclusive, paid-up, irrevocable, world-wide license to publish or reproduce the published form of this manuscript, or allow others to do so, for United States Government purposes. 

\section*{Declaration of Interests}
The authors report no conflict of interest.

\section*{Data availability statement}
The code used to generate the figures in this paper is available at \url{https://datacommons.princeton.edu/discovery/catalog/doi-10-34770-2zvn-wp92}. 
  
\appendix

\section{Fixed-$\theta$ variables for electrons}\label{Appendix fixed theta}
The concept of fixed-$\theta$ variables was first introduced by \cite{Trinczek23} for ions. Here, we extend the derivation to electrons.\par
Particles on trapped and passing orbits undergo changes in their parallel velocity $v_\parallel$ and radial position $\psi$ due to the magnetic and electric fields. In other words, the parallel velocity and radial position of a particle depend on the poloidal angle. The orbits are periodic and one can choose a reference angle $\theta_f$ to take the poloidal velocity and radial position at this reference angle as a constant of the motion of the particle. For passing particles, the choice of velocity and position at $\theta_f$ is unambiguous. Trapped particle orbits generally do not extend to all poloidal angles and cross each poloidal point once on their upwards leg and once on the downwards leg. Thus, the choice of the reference point is not unique. We capture all trapped particles and avoid double-counting by choosing $\vp=v_\parallel(\theta=\theta_f)$ positive for trapped particles and by first setting $\theta_f=0$, then $\theta_f=\pi$. This way, we can capture particles trapped on the outboard side, and particles that are potentially trapped by the poloidal variation of the correction to the electrostatic potential $\phi_{1}$ on the inboard side. \par 
The fixed-$\theta$ variables $\vp\equiv v_\parallel(\theta=\theta_f)$ and $\psi_f=\psi(\theta=\theta_f)$ together with the magnetic moment $\mu$ can be interpreted as labels of an orbit. If $\vp$, $\psi_f$ and $\mu$ are known, the corresponding orbit is uniquely determined. This formalism is equivalent to using the conserved quantities energy $\mathcal{E}\equiv v^2/2+Ze\Phi$, canonical angular momentum $\psi_\ast\equiv \psi-Iv_\parallel/\Omega$ and magnetic moment $\mu$, but the fixed-$\theta$ variables have the advantage that their deviation from the particle quantities $v_\parallel$ and $\psi$ is small in a large aspect ratio tokamak as will be shown in what follows.\par
The derivation of the orbit relations uses the conservation of energy and angular momentum. First, we expand in the smallness of the square root of the mass ratio $\delta$, keeping $\epsilon\sim 1$. The quantity $\psi-\psi_f$ is small in $\delta$ while $v_\parallel-\vp$ is small in $\epsilon$. Angular momentum conservation gives
\begin{equation}\label{angular step 0}
    \psi-\frac{I v_\parallel}{\Omega_e}=\psi_f-\frac{I v_{\parallel f}}{\Omega_{ef}}.
\end{equation}
The deviation from $\vp$ and $\psi_f$ as the particle completes its orbit are thus related via 
\begin{equation}\label{psi to v}
    \psi-\psi_f\simeq \frac{I}{\Omega_{ef}}(v_\parallel -v_{\parallel f})+\frac{I \vp}{\Omega_{ef}}\left(\frac{B_f}{B}-1\right)\sim\delta R B_p \rho_p.
\end{equation}
The energy of a particle is conserved, so
\begin{equation}\label{energy}
    \frac{v_\parallel^2}{2}+\mu B-\frac{e}{m_e}\Phi(\psi,\theta)=\frac{\vp^2}{2}+\mu_f B_f -\frac{e}{m_e}\Phi(\psi_f,\theta_f),
\end{equation}
which can be written as
\begin{equation}\label{energy step 2}
    \frac{(v_\parallel-\vp)^2}{2}+\vp(v_\parallel-\vp)+\mu B_f\left(\frac{B}{B_f}-1\right)=\frac{e}{m_e}\left[\Phi(\psi,\theta)-\Phi(\psi_f,\theta_f)\right].
\end{equation}
The electric potential $\Phi$ has a piece $\phi$ that is a flux function and a piece $\phi_1$ that varies with poloidal angle,
\begin{equation}
    \Phi(\psi,\theta)=\phi(\psi)+\phi_1(\psi,\theta),
\end{equation}
with $e\phi/T_e\sim1$ and $e\phi_1/T_e\sim\epsilon$. The electric potential can be expanded around $\psi_f$ as
\begin{equation}\label{expansion phi}
    \Phi(\psi,\theta)=\Phi(\psi_f,\theta)+(\psi-\psi_f)\pdv{\Phi}{\psi_f}+....
\end{equation}
From \eqref{psi to v} and the gradient lengthscale ordering $L_{\Phi}\sim\rho_p$, it follows that the second term of \eqref{expansion phi} is of order $\delta\Phi$ and thus small in $\delta$. The right hand side of \eqref{energy step 2} becomes
\begin{equation}
    \frac{e}{m_e}\left[\Phi(\psi,\theta)-\Phi(\psi_f,\theta_f)\right]= \frac{e}{m_e}\left[\phi_1(\psi_f,\theta)-\phi_1(\psi_f,\theta_f)  \right] +\textit{O}\left(\delta v_{te}^2\right).  
\end{equation}
\par We now expand in $\epsilon\ll 1$ for which we need to distinguish between trapped--barely passing and freely passing particles. Trapped--barely passing particles have an orbit width of $\sqrt{\epsilon}\delta \rho_p$ whereas freely passing particles have an orbit width of $\epsilon \delta \rho_p$. For freely passing particles, the first term in \eqref{energy step 2} can be dropped as small in $\epsilon$, such that the orbit equations for the passing particles read
\begin{equation}
    v_\parallel-\vp=-\frac{\mu B_f\left(B/B_f-1\right)+e\left[\phi_1(\psi_f,\theta)-\phi_1(\psi_f,\theta_f)\right]/m_e}{\vp}\sim \epsilon v_{te}
\end{equation}
and
\begin{equation}
    \psi-\psi_f=-\frac{I}{\Omega_{ef}}\frac{\left(\vp^2+\mu B_f\right)\left(B/B_f-1\right)+e\left[\phi_1(\psi_f,\theta)-\phi_1(\psi_f,\theta_f)\right]/m_e}{\vp}\sim \epsilon I\rho_e.
\end{equation}
For trapped--barely passing particles, all terms in \eqref{energy step 2} need to be kept because $\vp\sim \sqrt{\epsilon}v_{te}$, but the second term on the right of \eqref{psi to v} can be dropped as small in $\epsilon$ such that to lowest order in $\delta$, the energy equation becomes
\begin{equation}
    v_\parallel-\vp=-\vp +\sigma \sqrt{\vp^2-2\left[\mu B_f\left(\frac{B}{B_f}-1\right)-\frac{e}{m_e}\left(\phi_1(\psi_f,\theta)-\phi_1(\psi_f,\theta_f)\right) \right]}
\end{equation}
and 
\begin{equation}\label{psi to v trap}
    \psi-\psi_f=\frac{I}{\Omega_{ef}}(v_\parallel-\vp),
\end{equation}
where $\sigma=v_\parallel/\abs{v_\parallel}$ is the sign of the particle's parallel velocity.

\section{Lowest order distribution function for electrons}\label{Appendix lowest g}
The calculation of the lowest order distribution function for electrons in the trapped--barely passing region follows the ion calculation by \cite{Trinczek23}. The difference in the derivation is that the first step is an expansion in $\delta$ before a further expansion in $\sqrt{\epsilon}$ gives the final answer. The idea is to calculate $g_e$ from \eqref{fe} and \eqref{fe1}, 
\begin{equation}\label{simple g}
    g_e(v_\parallel,\psi)=f_{Mef}-f_{Me}+f_{e1f}(\vp,\psi_f)-\frac{m_ev_\parallel V_\parallel}{T_{e0}}f_{Me}.
\end{equation}
We use equation \eqref{energy} to write the Maxwellian as
\begin{multline}\label{fMe2}
    f_{Me}=n_{e0}(\psi)\left(\frac{m_e}{2\pi T_{e0}(\psi)}\right)^{3/2} \exp\bigg\lbrace -\frac{m_e\vp^2}{2 T_{e0}(\psi)}-\frac{m_e\mu B(\theta_f)}{T_{e0}(\psi)}+\frac{e\phi_1(\psi_f,\theta_f)}{T_{e0}(\psi)}\\
    -\frac{e}{T_{e0}(\psi)}\left[\phi(\psi)-\phi(\psi_f) \right]    \bigg\rbrace.
\end{multline}
We expand the density, temperature and electric potential in equation \eqref{fMe2} in $\psi-\psi_f$ in the same way as in \eqref{expansion phi}. Keeping terms up to order $\delta$, we find
\begin{multline}\label{fme to fmef}
    f_{Me}\simeq f_{Mef}+\bigg\lbrace \pdv{}{\psi_f}\ln p_{e0f} +\left[\frac{m_e \vp^2}{2T_{e0f}}+\frac{m_e\mu B_f}{T_{e0f}}-\frac{5}{2}\right]\pdv{}{\psi_f}\ln T_{e0f}\\
   +\frac{m_e u_f}{T_{e0f}}\frac{\Omega_{ef}}{I}\bigg\rbrace (\psi-\psi_f) f_{Mef},
\end{multline}
where $T_{e0f}\equiv T_{e0}(\psi_f)$ and $n_{e0f}\equiv n_{e0}(\psi_f)$.
For trapped--barely passing particles, $v_\parallel\sim \sqrt{\epsilon}v_{te}$, so 
\begin{equation} \label{ge simple 2}
    g_e(v_\parallel,\psi)\simeq f_{Mef}-f_{Me}+g_{ef}(\vp,\psi_f)-(v_\parallel-\vp)\frac{m_e V_{\parallel f}}{T_{e0}(\psi_f)}f_{Mef},
\end{equation}
where $g_{ef}(\vp,\psi_f)=f_{e1f}(\vp,\psi_f)-m_e\vp V_{\parallel f}/T_{e0}(\psi_f)$.
To lowest order in $\sqrt{\epsilon}$, \eqref{psi to v trap} holds and thus combining \eqref{fme to fmef} and \eqref{ge simple 2} gives
\begin{equation}
    g_e=-(v_\parallel-\vp)\alpha_{0e}+g_{ef}(\vp,\psi_f),
\end{equation}
where $\alpha_{0e}$ was defined in \eqref{alpha0e}. In the definition of $\alpha_{0e}$, all $f$ and $0$ subscripts were dropped where possible to simplify the notation.
The function $g_{ef}$ can be solved for using the procedure presented by \cite{Trinczek23} and gives the results for $g_{e0}$ in \eqref{dg0e t}. The largest piece of $g_e$ is $g_{e0}\sim \sqrt{ \epsilon}\delta f_{Me}$, and there is no $g_{e}\sim \delta f_{Me}$ piece. 

\section{Alternative calculation of the bootstrap current}\label{Appendix bootstrap}
When the bootstrap current was derived in section \ref{sec:bootstrap}, we used self-adjointness of the collision operator in \eqref{self adjoint}. However, one can also switch to fixed-$\theta$ variables and use the jump properties of the distribution function to take the integrals and derive the same expression for the bootstrap current. For this alternative derivation that treats the discontinuities more carefully, we substitute the expression for $v_\parallel$ in terms of the Spitzer-Härm function \eqref{Spitzer vparallel} into \eqref{jb step 2} which gives 
\begin{multline}\label{1st IBP}
    \langle j^B_\parallel\rangle_\psi=-e\bigg\langle \int_{V_p}\mathrm{d}^3v_f \:  \frac{g_e}{f_{Me}}\bnabla_v\bcdot\bigg[f_{Me}\bm{\mathsf{{M_{e}}}}\bcdot 
\bnabla_v\left(\frac{f_{e,SH}}{f_{Me}}\right)\\
-\lambda_e f_{Me}\int\mathrm{d}^3v'\: f_{Me}'\bnabla_\omega\bnabla_\omega\omega \bcdot \bnabla_{v'}\left(\frac{f_{e,SH}'}{f'_{Me}}\right) \bigg]\bigg\rangle_\psi\\
=e\Bigg\langle \int\mathrm{d}\mu\: 2\pi B_f \Delta\bigg[g_e \boldsymbol{\hat{b}}\bcdot \bm{\mathsf{{M_e}}}\bcdot 
\bnabla_v\left(\frac{f_{e,SH}}{f_{Me}}\right)\\
-\lambda_e g_e\int\mathrm{d}^3v'\: f_{Me}'\bnabla_\omega\bnabla_\omega\omega \bcdot \bnabla_{v'}\left(\frac{f_{e,SH}'}{f'_{Me}}\right) \bigg]\Bigg\rangle_\psi\\
+e\Bigg\langle \int_{V_p}\mathrm{d}^3v_f \bnabla_v\left(\frac{g_e}{f_{Me}}\right)\bcdot\bigg[f_{Me}\bm{\mathsf{{M_e}}}\bcdot 
\bnabla_v\left(\frac{f_{e,SH}}{f_{Me}}\right)\\
-\lambda_e f_{Me}\int\mathrm{d}^3v'\: f_{Me}'\bnabla_\omega\bnabla_\omega\omega \bcdot \bnabla_{v'}\left(\frac{f_{e,SH}'}{f'_{Me}}\right) \bigg]\Bigg\rangle_\psi.
\end{multline}
We integrated by parts and picked up the jump at the trapped--barely passing region $\vp=-u_f\simeq0$. In what follows, we rewrite each of these terms in a convenient form.\par 
The gradient of the Spitzer-Härm function is 
\begin{equation}
    \bnabla_v\left(\frac{f_{e,SH}}{f_{Me}} \right)=\frac{1}{\sqrt{2}\nu_{ee}}\left[\boldsymbol{\hat{b}}A_{SH}+v_\parallel\frac{\bm{v}}{v_{te}v}\pdv{A_{SH}}{x_e}\right],
\end{equation}
such that to lowest order the jump terms become 
\begin{multline}\label{first jump j}
    e\Bigg\langle \int\mathrm{d}\mu\: 2\pi B_f \Delta\bigg[g_e \boldsymbol{\hat{b}}\bcdot \bm{\mathsf{{M_e}}}\bcdot 
\bnabla_v\left(\frac{f_{e,SH}}{f_{Me}}\right)\\
-\lambda_e g_{e}\int\mathrm{d}^3v'\: f_{Me}'\bnabla_\omega\bnabla_\omega\omega \bcdot \bnabla_{v'}\left(\frac{f_{e,SH}'}{f'_{Me}}\right) \bigg]\Bigg\rangle_\psi\\
=\Bigg\langle \int\mathrm{d}\mu\: 2\pi B_f\Delta g_e\bigg[ \frac{e}{\sqrt{2}\nu_{ee}}\mathsf{M_{\parallel e}} A_{SH}-e\lambda_e \int\mathrm{d}^3v'\: f_{Me}'\bnabla_\omega\bnabla_\omega\omega \bcdot \bnabla_{v'}\left(\frac{f_{e,SH}'}{f'_{Me}}\right) \bigg]_{v_\parallel=0}\Bigg\rangle_\psi.
\end{multline}
In this expression, the functions $ \mathsf{M_{\parallel e}}$ and $A_{SH}$, the Maxwellian $f_{Me}$ and $\bm{\omega}=\bm{v}-\bm{v'}$ are evaluated at $v_\parallel=0$. To lowest order, the only quantity in this expression experiencing a jump is the electron distribution function. The jump is given in \eqref{wf int}.

Next, we integrate by parts the third term in \eqref{1st IBP} containing the derivative of $g_e$, 
\begin{multline}\label{2nd IBP}
e\Bigg\langle \int_{V_p}\mathrm{d}^3v_f \bnabla_v\left(\frac{g_e}{f_{Me}}\right)\bcdot \bm{\mathsf{{M_e}}}\bcdot 
\bnabla_v\left(\frac{f_{e,SH}}{f_{Me}}\right)f_{Me}\Bigg\rangle_\psi\\
= -e\Bigg\langle\int_{V_p}\mathrm{d}^3 v_f\:\frac{f_{e,SH}}{f_{Me}}\bnabla_v\bcdot\left[ f_{Me}\bm{\mathsf{M_e}}\bcdot \bnabla_v\left(\frac{g_e}{f_{Me}}\right)\right]\Bigg\rangle_\psi\\
-e\Bigg\langle \int\mathrm{d}\mu\: 2\pi B_f \Delta\left[f_{e,SH} \boldsymbol{\hat{b}} \bcdot\bm{\mathsf{M_e}}\bcdot \bnabla_v\left(\frac{g_e}{f_{Me}}\right)\right]\Bigg\rangle_\psi.
\end{multline}
The second term on the right in \eqref{2nd IBP} vanishes because the integrand is evaluated in the limit $v_\parallel\rightarrow 0^{+,-}$, where $f_{e,SH}=0$.\par
For the last term in \eqref{1st IBP}, we can write 
\begin{multline}
    -e\Bigg\langle\int_{V_p}\mathrm{d}^3v_f\int\mathrm{d}^3v' \lambda_e f_{Me}f'_{Me}\bnabla_v\left(\frac{g_e}{f_{Me}}\right)\bcdot\bnabla_\omega\bnabla_\omega\omega\bcdot\bnabla_{v'}\left(\frac{f_{e,SH}'}{f'_{Me}	}\right)\Bigg\rangle_\psi\\
    \simeq -e\Bigg\langle\int\mathrm{d}^3v\int\mathrm{d}^3v' \lambda_ef_{Me}f'_{Me}\bnabla_v\left(\frac{g_e}{f_{Me}}\right)\bcdot\bnabla_\omega\bnabla_\omega\omega\bcdot\bnabla_{v'}\left(\frac{f_{e,SH}'}{f'_{Me}	}\right)\Bigg\rangle_\psi\\
    +e\Bigg\langle\int_{V_{t,bp}}\mathrm{d}^3v\int\mathrm{d}^3v' \lambda_e f_{Me}f'_{Me}\bnabla_v\left(\frac{g_e}{f_{Me}}\right)\bcdot\bnabla_\omega\bnabla_\omega\omega\bcdot\bnabla_{v'}\left(\frac{f_{e,SH}'}{f'_{Me}	}\right)\Bigg\rangle_\psi. 
\end{multline}
Exchanging $\bm{v}$ and $\bm{v'}$ in the first term and taking the $v_\parallel$ integral over the trapped region in the second term, we get 
\begin{multline}\label{Integral terms j}
     -e\Bigg\langle\int_{V_p}\mathrm{d}^3v_f\int\mathrm{d}^3v' \lambda_e f_{Me}f'_{Me}\bnabla_v\left(\frac{g_e}{f_{Me}}\right)\bcdot\bnabla_\omega\bnabla_\omega\omega\bcdot\bnabla_{v'}\left(\frac{f_{e,SH}'}{f'_{Me}	}\right)\Bigg\rangle_\psi\\
    \simeq- e\Bigg\langle \int\mathrm{d}^3v\: \lambda_e f_{Me} \bnabla_v\left(\frac{f_{e,SH}}{f_{Me}}\right)\bcdot \int\mathrm{d}^3v'\: f_{Me}'\bnabla_\omega\bnabla_\omega\omega\bcdot\bnabla_{v'}\left(\frac{g_e'}{f'_{Me}	}\right)\bigg\rangle_\psi\\
    +e\Bigg\langle \int\mathrm{d}\mu\: 2\pi B_f \Delta g_e \lambda_e \int\mathrm{d}^3v'\:  f_{Me}'\bnabla_\omega\bnabla_\omega\omega\bcdot\bnabla_{v'}\left(\frac{f_{e,SH}'}{f'_{Me}}\right)\Bigg\rangle_\psi.
\end{multline}
The last term in \eqref{Integral terms j} cancels the second term in \eqref{first jump j}. Combining \eqref{1st IBP}, \eqref{first jump j}, \eqref{2nd IBP} and \eqref{Integral terms j} gives 
\begin{multline}
     \langle j^B_\parallel\rangle_\psi=\Bigg\langle \int\mathrm{d}\mu\: 2\pi B \Delta g_e \frac{e}{\sqrt{2}\nu_{ee}} \mathsf{M_{\parallel e}}A_{SH}\Big\rvert_{v_\parallel=0}\Bigg\rangle_\psi\\
      -e\Bigg\langle\int_{V_p}\mathrm{d}^3 v_f\:\frac{f_{e,SH}}{f_{Me}}\bnabla_v\bcdot\left[f_{Me} \bm{\mathsf{M_e}}\bcdot \bnabla_v\left(\frac{g_e}{f_{Me}}\right)\right]\Bigg\rangle_\psi \\
      -e\Bigg\langle \int\mathrm{d}^3v\: \lambda_e f_{Me} \bnabla_v\left(\frac{f_{e,SH}}{f_{Me}}\right)\bcdot \int\mathrm{d}^3v'\: f_{Me}'\bnabla_\omega\bnabla_\omega\omega\bcdot\bnabla_{v'}\left(\frac{g_e'}{f'_{Me}	}\right)\bigg\rangle_\psi.
\end{multline}
When the last term is integrated by parts again, the combination of the second and the third term gives the collision operator $C_e$, such that 
\begin{multline}\label{j as jump}
     \langle j^B_\parallel\rangle_\psi=\Bigg\langle \int\mathrm{d}\mu\: 2\pi B \Delta g_e \frac{e}{\sqrt{2}\nu_{ee}} \mathsf{M_{\parallel e}}A_{SH}\Big\rvert_{v_\parallel=0}\Bigg\rangle_\psi-e\Bigg\langle \int_{V_p} \mathrm{d}^3v_f \frac{f_{e,SH}}{f_{Me}}C_e\Bigg\rangle_\psi,
\end{multline}
where $\mathsf{M_{\parallel e}}$ and $A_{SH}$ are evaluated at $v_\parallel=0$, e.g. $x_e^2=m_e\mu B/T_e$. 
The flux surface average and the transit average for freely passing particles are equivalent, such that we can use the expression for the drift kinetic equation \eqref{drift kinetic eq} to argue that the last term of \eqref{j as jump} vanishes. Indeed, the source term \eqref{sigma} is small, so this term is of order $\delta^2\sqrt{\epsilon} e n_ev_{te}$ and can be neglected. Using the jump condition \eqref{wf int}, we arrive back at the expression for the bootstrap current in \eqref{bootstrap integral}.

\section{Input profiles for case study}\label{sec: Input formulas}
The profiles for density, ion and electron temperature are based on those measured by \cite{Viezzer17}. We use the analytical expressions
\begin{equation}
\bar{n}=0.6035+0.3965\tanh\left[-1.2929(\bar{\psi}-9.3942)\right]-0.0075\bar{\psi},
\end{equation}
\begin{equation}
    \bar{T}_i=1-0.0459\bar{\psi}+0.0038\bar{\psi}^2-0.0007\bar{\psi}^3,
\end{equation}
and
\begin{equation}
    \bar{T}_e=1.2648-0.2798\tanh\left[1.3578(\bar{\psi}-9.0470) \right]-0.0871\bar{\psi}.
\end{equation}

\section{Poloidal asymmetry effects}\label{Appendix poloidal}
There are four sources of poloidal asymmetry - passing particle number asymmetry, centrifugal force, mean parallel flow gradient, and orbit width asymmetry. We split up the terms in \eqref{phic} and attribute them to these four effects in figure \ref{fig: phicterms}. The categorization is not unique as there are many crossterms between the four effects. However, for the purpose of figure \ref{fig: phicterms}, we use
\begin{multline}
    \text{number asymmetry}=-Z\sqrt{\bar{T}}\bigg\lbrace  J\left( \frac{2\bar{u}^2}{\bar{T}}+1\right)\bigg[\pdv{}{\bar{\psi}}\ln \bar{p}-\frac{3}{2}\pdv{}{\bar{\psi}}\ln \bar{T}\\
    -\frac{2(\bar{u}+\bar{V})}{\bar{T}}\left(\pdv{\bar{V}}{\bar{\psi}}-1\right)\bigg]  +\pdv{}{\bar{\psi}}\ln \bar{T}\left[J-\frac{\bar{u}+\bar{V}}{2\sqrt{\bar{T}}}\left(1-2\frac{\bar{u}+\bar{V}}{\sqrt{\bar{T}}}J\right)\left(\frac{2\bar{u}^2}{\bar{T}}+1\right)  \right]\bigg\rbrace/\mathcal{N},
\end{multline}
\begin{equation}
    \text{centrifugal force}=\left(\frac{2\bar{u}^2}{\bar{T}}\right)/\mathcal{N},
\end{equation}
\begin{equation}
    \text{flow gradient}=-2\pdv{\bar{V}}{\bar{\psi}}\left(1+\frac{\bar{u}^2}{\bar{T}}\right)/\mathcal{N},
\end{equation}
\begin{equation}
    \text{orbit width}= \left[-(\bar{V}-\bar{u})\pdv{}{\bar{\psi}}\ln\bar{n}\right]/\mathcal{N},
\end{equation}
where
\begin{multline}
    \mathcal{N}=\frac{1}{\bar{T}_e}-\frac{Z}{\bar{T}}\bigg[J\sqrt{\bar{T}}\left(\pdv{}{\bar{\psi}}\ln\bar{p}-\frac{3}{2}\pdv{}{\bar{\psi}}\ln\bar{T}\right)\\
    +\left(1-2\frac{\bar{u}+\bar{V}}{\sqrt{\bar{T}}}J\right)\left(\pdv{\bar{V}}{\bar{\psi}}-1-\frac{\bar{V}+\bar{u}}{2}\pdv{}{\bar{\psi}}\ln\bar{T}\right)\bigg].
\end{multline}

\section{Weak gradient limit}\label{sec:weak comp}
Our results for the neoclassical particle flux and the bootstrap current reduce to the standard neoclassical weak gradient results for high flow and low flow in the appropriate limit.\par 
In the limit of weak gradients, the high flow ion particle flux equation gives 
\begin{equation}\label{V+u}
    V_{\parallel }+u=-\frac{I T_{i}}{m_i\Omega_{i}}\left(\pdv{}{\psi}\ln p_{i}-1.17\frac{G_2(0,\bar{z}^{wg,hf})}{G_1(0,\bar{z}^{wg,hf})}\pdv{}{\psi}\ln T_{i}\right),
\end{equation}
where $G_1$ and $G_2$ are given in (5.13) and (5.14) in \cite{Trinczek23} and $\bar{z}^{wg,hf}=mu^2/T-Ze\phi_c^{wg,hf}R/Tr$. The first argument of $G_1$ and $G_2$ vanishes because $V_\parallel+u\simeq 0$.
In the high flow limit, centrifugal forces drive the poloidal potential variation 
\begin{equation}\label{phi centri}
    \left(\frac{en_e}{T_e}+\frac{Z^2en_i}{T_i}\right)\phi_c^{wg,hf}=Zn_i\frac{r}{R}\frac{m_iu^2}{T_i}.
\end{equation}
Using both expressions in \eqref{Gammae} gives the weak gradient-high flow neoclassical electron particle flux 
\begin{multline}
    \Gamma_e^{wg,hf}\simeq-3.17\frac{\nu_{ee} I^2 p_e}{\Omega_e^2 m_e}\sqrt{\frac{r}{R}}\Bigg\lbrace  \bigg[\left(1+\frac{T_{i}}{ZT_{e}}\right)\pdv{}{\psi}\ln n_{e}  \\
    + \frac{T_{i}}{ZT_{e}}\left(1-1.17\frac{G_2(0,\bar{z}^{wg,hf})}{G_1(0,\bar{z}^{wg,hf})} \right)\pdv{}{\psi}\ln T_{i}   \bigg] G_{1e}(\phi_c^{wg,hf},Z)\\
    +\left[G_{1e}(\phi_c^{wg,hf},Z)-1.39 G_{2e}(\phi_c^{wg,hf},Z)\right]\pdv{}{\psi}\ln T_{e}\Bigg\rbrace.
\end{multline}
Similarly, the weak gradient-high flow limit of the bootstrap current \eqref{jb} is
\begin{multline}
    \langle j_\parallel^B\rangle^{wg,hf}\simeq-2.43 \frac{cI}{B}\sqrt{\frac{r}{R}}p_e\Bigg\lbrace  \bigg[\left(1+\frac{T_{i}}{ZT_{e}}\right)\pdv{}{\psi}\ln n_{e} \\
    + \frac{T_{i}}{ZT_{e}}\pdv{}{\psi}\ln T_{i}\left(1-1.17\frac{G_2(0,\bar{z}^{wg,hf})}{G_1(0,\bar{z}^{wg,hf})}\right)  \bigg]J_{1e}(\phi_c^{wg,hf},Z)  \\
    +\left[J_{1e}(\phi_c^{wg,hf},Z)-0.71 J_{2e}(\phi_c^{wg,hf},Z)\right]\pdv{}{\psi}\ln T_{e}\Bigg\rbrace.
\end{multline}
\par In the low flow limit, the centrifugal force is weak and the poloidal variation of the electric potential vanishes, $\phi_c\rightarrow0$. For $Z=1$, the particle flux in the weak gradient limit reduces to the known result for the low flow regime
\begin{equation}\label{Gammae wg}
    \Gamma_e^{wg,lf}=-3.17\frac{\nu_{ee} I^2 p_{e}}{\Omega_{e}^2 m_e}\sqrt{\frac{r}{R}}  \left[\left(1+\frac{T_{i}}{T_{e}}\right)\pdv{}{\psi}\ln n_{e} -\frac{0.17}{T_{e}}\pdv{T_{i}}{\psi}-0.39 \pdv{}{\psi}\ln T_{e}\right].
\end{equation}
Similarly, the expression for the bootstrap current \eqref{jb} in the weak gradient-low flow limit and $Z=1$ reduces to
\begin{equation}\label{jb wg}
    \langle j_\parallel ^B\rangle_\psi^{wg,lf}=-2.43 \frac{cI}{B_f}\sqrt{\frac{r}{R}}p_{e}\left[\left(1+\frac{T_{i}}{T_{e}}\right)\pdv{}{\psi}\ln n_{e} -\frac{0.17}{T_{e}}\pdv{T_{i}}{\psi}+0.29 \pdv{}{\psi}\ln T_{e} \right],
\end{equation}
which is in agreement with the weak gradient-low flow neoclassical prediction \citep{helander05}.\par


\bibliographystyle{jpp}
\bibliography{BootstrapBib}

\end{document}